%% file: JSS_2019.tex
\documentclass[review]{elsarticle}

\AtBeginDocument{}

\usepackage{textcomp}
\usepackage{graphicx}
\usepackage{subfig}

\usepackage[table]{xcolor}

\usepackage{csquotes}

\usepackage{tikz}
\usetikzlibrary{fit} 
\usetikzlibrary{positioning}
\usetikzlibrary{arrows}
\usetikzlibrary{shapes.multipart}

\usepackage[hidelinks,bookmarks=false]{hyperref}
\usepackage[numbered]{bookmark}
\usepackage{soul}
\usepackage{booktabs}
\usepackage{multirow}
\usepackage{float}
\usepackage{url}
\usepackage{tcolorbox}
\usepackage{amsmath,amssymb,amsfonts}
\usepackage{algorithmic}
\usepackage{textcomp}
\usepackage{url}
\usepackage{longtable}
\journal{Journal of \LaTeX\ Templates}

\usepackage{pgfplots}
\pgfplotsset{width=7cm,compat=1.8,tick label style={font=\small}}

\usepackage{adjustbox}
\usepackage{float}
\usepackage{rotating}
\usepackage{tablefootnote}
\usepackage{nth}
\DeclareCaptionType{TextBox}
\newcolumntype{L}{>{\centering\arraybackslash}m{3cm}}

\usepackage[utf8]{inputenc}
\usepackage{dirtytalk}
\usepackage{tcolorbox}

\usepackage[british]{babel}
\usepackage{enumitem}

\newlist{SubItemList}{itemize}{1}
\setlist[SubItemList]{label={$-$}}

\let\OldItem\item
\newcommand{\SubItemStart}[1]{%
    \let\item\SubItemEnd
    \begin{SubItemList}[resume]%
        \OldItem #1%
}
\newcommand{\SubItemMiddle}[1]{%
    \OldItem #1%
}
\newcommand{\SubItemEnd}[1]{%
    \end{SubItemList}%
    \let\item\OldItem
    \item #1%
}
\newcommand*{\SubItem}[1]{%
    \let\SubItem\SubItemMiddle%
    \SubItemStart{#1}%
}

\begin{document}





\begin{frontmatter}

\title{Toward the Automatic Classification of Self-Affirmed Refactoring}

\author[RIT]{Eman Abdullah AlOmar\corref{mycorrespondingauthor}}
\cortext[mycorrespondingauthor]{Corresponding author}
\ead{eman.alomar@mail.rit.edu}

\author[RIT]{Mohamed Wiem Mkaouer}
\ead{mwmvse@rit.edu}

\author[ETS]{Ali Ouni}
\ead{ali.ouni@etsmtl.ca}

\address[RIT]{Rochester Institute of Technology, Rochester, NY, USA}
\address[ETS]{ETS Montreal, University of Quebec, Montreal, QC, Canada}

\begin{abstract}
The concept of Self-Affirmed Refactoring (SAR) was introduced to explore how developers document their refactoring activities in commit messages, \textit{i.e.,} 
developers explicit documentation of refactoring operations intentionally introduced during a code change. In our previous study, we have manually identified refactoring patterns and defined three main common quality improvement categories 
including internal quality attributes, external quality attributes, and code smells, by only considering refactoring-related commits. However, this approach heavily depends on the manual inspection of commit messages. In this paper, we propose a two-step approach to first identify whether a commit describes developer-related refactoring events, 
 then to classify it according to the refactoring common quality improvement categories. 
   Specifically, we combine the N-Gram TF-IDF feature selection with binary and multiclass classifiers to build a new model to automate the classification of refactorings based on their quality improvement categories. We challenge our model using a total of 2,867 commit messages extracted from well engineered open-source Java projects. Our findings show that (1) our model is able to accurately classify SAR commits, outperforming the pattern-based and random classifier approaches, and allowing the discovery of 40 more relevant SAR patterns, 
   and (2) our model reaches an F-measure of up to 90\% even with a relatively small training dataset.
\end{abstract}


\end{frontmatter}


\input{Sections/Introduction.tex}
\input{Sections/self-affirmed}
\input{Sections/RelatedWork.tex}

\input{Sections/Approach.tex}

\input{Sections/Results.tex}
\input{Sections/Implications.tex}
\input{Sections/Threats.tex}
\input{Sections/Conclusion.tex}

\bibliography{references}

\end{document}

%% file: Sections/Introduction.tex
\section{Introduction}
\label{sec:Introduction}


The role of refactoring has been growing from simply improving the internal structure of the code without altering its external behavior \cite{Fowler:1999:RID:311424} to hold a key driver of the agile methodologies and become one of the main practices to reduce technical debt \cite{fowler2001agile}. According to recent surveys, the research on refactoring has been focused on automating it through recommending candidate code elements to be refactored and which refactoring operations to apply \cite{tsantalis2008jdeodorant,bavota2013empirical,bavota2014recommending,charalampidou2016identifying}. Yet, more recent studies have shown that fully automated techniques are underused in practice \cite{kim2014empirical}. Indeed, there is a need to minimize the disturbance of the existing design, by performing large refactorings, as developers typically want to recognize and preserve the semantics of their own design, even at the expense of not significantly 
improving it \cite{kim2014empirical,ouni2016multi,ouni2012search}.

Therefore, several studies have taken a developer-centric approach by detecting how developers do refactor their code \cite{Silva:2016:WWR:2950290.2950305,silva2017refdiff,tsantalis2018accurate} and how they document their refactoring strategies \cite{liang2018,alomar2019can}. The detection of refactoring operations and their documentation allows a better understanding of code evolution, and challenges that trigger refactoring, including the reduction of code proneness to errors, facilitation of API and type migrations, etc. While automating the detection of refactoring operations that are applied in the source code has advanced recently reaching a high accuracy \cite{tsantalis2018accurate}, there is a critical need for a deeper analysis of how such refactoring activities are being documented. In this context, recent studies \cite{liang2018,alomar2019can} have introduced a taxonomy on how developers actually document their refactoring strategies in commit messages. Such documentation is known as \textit{Self-Admitted} or \textit{Self-Affirmed} refactoring. Documenting refactoring, similarly to any type of code change documentation, is useful to decipher the rationale behind any applied change, and it can help future developers in various engineering tasks, such as program comprehension, design reverse-engineering, and debugging. However, the detection of such refactoring documentation was hardly manual and limited. There is a need for automating the detection of such documentation activities, with an acceptable level of accuracy. 
Indeed, the automated detection of refactoring documentation may support various applications and provide actionable insights to software practitioners and researchers, including empirical studies around the developer’s perception of refactoring. This can question whether developers do care about structural metrics and code smells when refactoring their code, or if there are other factors that may influence such non-functional changes. Furthermore, our previous study \cite{alomar2019can} found that there are several intentions behind the application of refactoring, which can be classified as improving internal structural metrics (\textit{e.g.,} cohesion, encapsulation), removing code smells (\textit{e.g.,} God classes, dead code), or optimizing external quality attributes (\textit{e.g.,} testability, readability). Yet, there is no systematic way to classify such refactoring related messages and estimate the distribution of refactoring effort among these categories.

To cope with the above-mentioned limitations, this paper aims to automate the detection and classification of refactoring documentation in commit messages. In particular, our objective is to analyze the feasibility and performance of applying learning techniques to (1) identify and (2) classify refactoring documentation based on commit messages. However, the detection of refactoring documentation is challenging, besides the inherited ambiguity of distinguishing meanings, in any natural language text, a recent study has shown that developers do misuse the term \textit{refactoring} in their documentations \cite{zhangpreliminary18}, which hardens the reliance on that keyword alone. To cope with these challenges, we design our study to harvest a potential taxonomy that can be used to document refactoring activities. Such taxonomy is typically threatened by the potential false-positiveness of the collected samples. Therefore, we develop a baseline of code changes that are known to contain refactoring activities, and we analyze their commit messages, in order to ensure that the collected textual patterns are meant to describe refactoring, and so, to reduce false positives.
Our study makes the following contributions:

\begin{itemize}
\item We present a two-step approach that firstly distinguishes whether a commit message potentially contains an explicit description of a refactoring effort. Then, secondly classifies it into one of the three common categories identified in previous studies \cite{liang2018,alomar2019can}. To the best of our knowledge this is the first attempt to automate the detection and classification of self affirmed refactorings.

\item We evaluate the performance of our approach by comparing it against a keyword-based approach that relies on matching messages with known refactoring keywords  \cite{kim2014empirical,liang2018,Ratzinger:2008:RRS:1370750.1370759,stroggylos2007refactoring,citeulike:2881658,soares2013comparing}. Our key findings show that our model not only outperforms the keyword-based approach, but also accurately identifies refactoring related commits with an average accuracy of 98\% and F-measure of 98\%. Furthermore, we infer which features, \textit{i.e.,} keywords, are relevant for the detection of such refactoring documentation, and we extract them to extend the list of refactoring documentation patterns, identified in previous studies \cite{liang2018,alomar2019can}.

\item We deploy our model as a lightweight web-service that is publicly available for software engineers and practitioners. We also publicly provide our dataset that served us as the \textit{ground-truth}, for replication and extension purposes \cite{SAR2019WEB}.
\end{itemize}

This paper is structured as follows. We start by explaining the notion of refactoring documentation (self-affirmed refactoring) 
and reviewing existing studies that are most related to commit messages classification in Section \ref{sec:RelatedWork}. 
Next, in Section \ref{sec:Approach}, we detail our two-step classification methodology. More precisely, we elaborate on the data collection and preprocessing, choice of the classification algorithms. Then, we evaluate our approach, in Section \ref{sec:ResultsDiscussions}, and report a comparative study between various classifiers, extracted from previous studies, and we identify most influential features. We report in Section \ref{sec:Threats} the threats to our work's validity, before concluding
and describing our future work in Section \ref{sec:conclusion}.

%% file: Sections/self-affirmed.tex
\section{Self-Affirmed Refactoring}
\label{section:self}

\subsection{Definition}

Commit messages are the atomic descriptions of given code change, in natural language. It augments the change with human and machine readable meaning. In this study, we are interested in locating and automatically detecting developer's documentation of refactoring activities in commit messages. \textit{refactoring documentation} is the textual description of what developers considers to be a refactoring performed in their code change. The act of intentionally documenting a refactoring activity is known as \textit{Self-Affirmed Refactoring} (SAR) \cite{alomar2019can}. SAR is composed of a terminology that was found to be consistently used in refactoring-related commit messages. For example, if we consider the following commit message:

\begin{displayquote}
Refactor createOrUpdate method in MongoChannelStore to extract methods and make code more readable
\end{displayquote}

\input{Tables/Pattern.tex}

The developer explicitly mentions the intention of refactoring, using the keyword \say{\textit{refactor}}, along with providing extra information related to the refactoring activity: the developer reports 1) the type of refactoring operation performed, \textit{i.e.}, \textit{extract method}; 2) the code elements involved in the refactoring operation, \textit{i.e.}, \textit{createOrUpdate} and \textit{MongoChannelStore}; and 3) the intent behind the refactoring, \textit{i.e.}, \textit{make the code more readable}. This message is labeled as Self-Affirmed Refactoring (SAR) as it totally or partially documents the refactoring performed in the source code.

The manual inspection of the message's corresponding commit\footnote{https://github.com/atlasapi/atlas-persistence}, reveals 3 methods extracted from the method \textit{createOrUpdate()} that belongs to the class \textit{MongoChannelStore} along with renaming a parameter to be consistent with the update. So, the documentation has given enough background to explain the rationale behind the refactoring (improving code readability), the operations performed and the code elements involved.

\subsection{Categories}


In our previous work \cite{alomar2019can}, we manually analyzed commit messages to extract any relevant textual patterns that can be considered as SAR. We provided a set of 87 SAR patterns, identified across 3,795 open source projects. Table~\ref{Table:Patterns} demonstrates all of these patterns. Since refactoring research typically focus on the detection of refactoring opportunities in the source code to recommend appropriate operations, we were particularly interested in extracting the intent behind the refactoring, to capture what typically triggers developers to refactor their code. As seen in Table~\ref{Table:Patterns}, intents can be either 1) \textit{generic}, using high-level keywords, such as \textit{Code cleanup}, \textit{Code revision}, \textit{Code reformatting \& reordering} etc.; or 2) \textit{specific}, using keywords that are more in line with the concepts used by tools to recommend refactoring. To further ensure the correctness of our data, we conducted a pilot study with a sample of data to learn, explore, and understand what challenges we faced when classifying commit messages. Based on the pilot study, we define the three SAR categories (\textit{i.e.,} internal, external, and code smell). In particular, developers typically state structural, size, complexity, and Object-Oriented metrics, such as coupling, composition, design size, etc. These metrics are the main drivers for many refactoring techniques \cite{bavota2014recommending,du2004refactoring,moser2007case,singh2012evaluation,mkaouer2015many,silva2014recommending}, and they are known in literature as \textit{internal quality attributes}.

Also, developers do mention the correction and management of bad programming practices, also known in refactoring studies \cite{tsantalis2008jdeodorant,ouni2016multi,Silva:2016:WWR:2950290.2950305,moha2009decor,szoke2014bulk,cedrim2017understanding} as \textit{code smells}, \textit{anti-patterns}, and \textit{design defects}. Code Smell resolution is the removal of design defects that might violate the fundamentals of software design principles and decrease code quality. Examples of these code smells include duplicate code, dead code, long method, blob class, etc.

Finally, we extracted intents corresponding to what literature considers as \textit{external quality attributes}. External quality attribute is the property or feature that indicates the effectiveness of a system such as understandability and readability. Many refactoring approaches are driven by the optimization of non-functional attributes such as testability, understandability, changeability, evolvability, and readability \cite{du2005does,geppert2005refactoring,ratzinger2005improving,newman2018study,newmanabbrev,newman2020generation}.

\input{Tables/QACodeSmell.tex}

\begin{sidewaysfigure}[htbp]
\centering 
\includegraphics[width=1.15\textwidth]{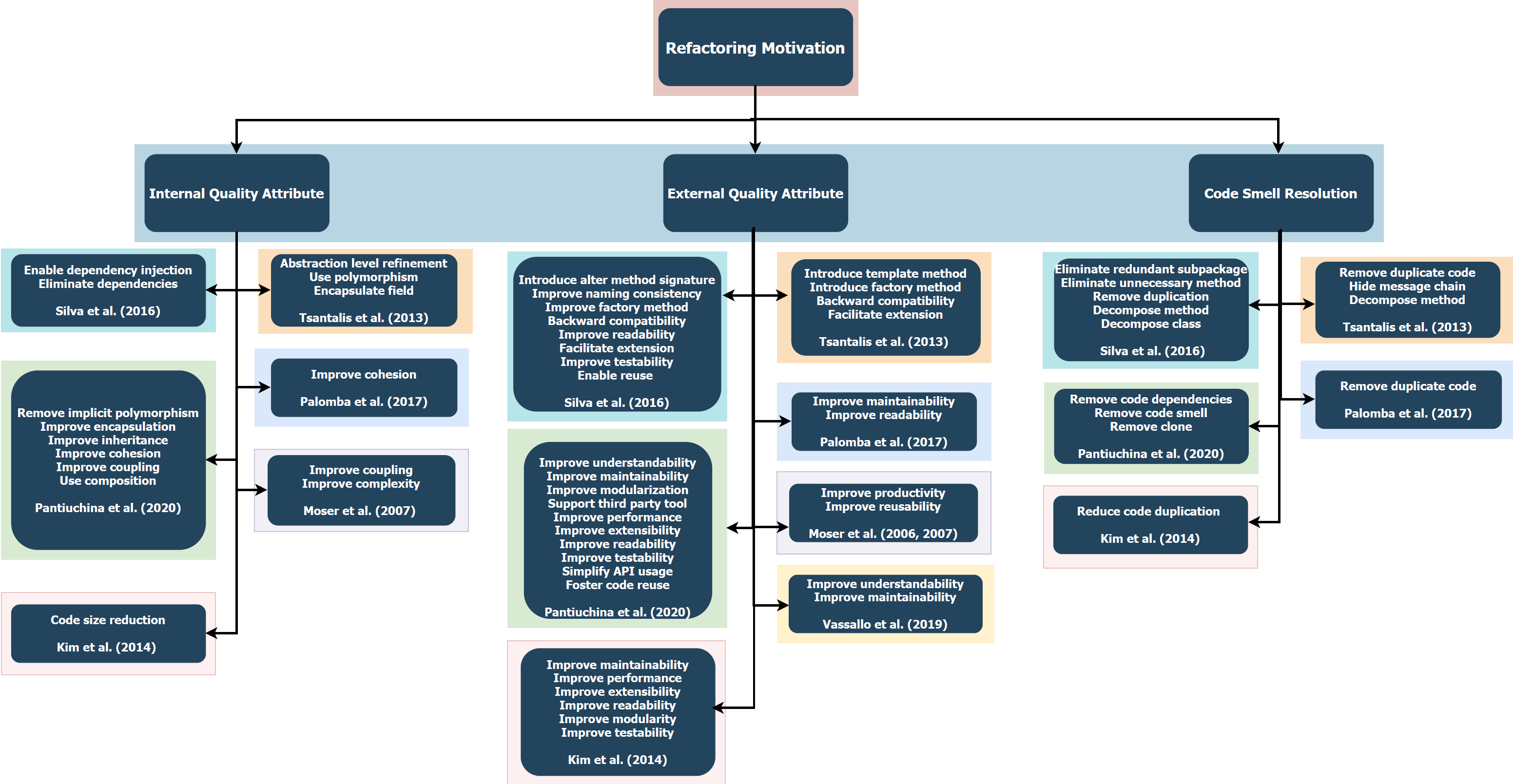}
\caption{Refactoring Motivation.}
\label{fig:motivation}
\end{sidewaysfigure}

The complete list of identified SAR patterns, per category, is depicted in Table~\ref{Table:QA & Code smell}. These categories, namely, \textit{internal quality attributes}, \textit{code smells}, and \textit{external quality attributes} represent what existing refactoring techniques are using to identify refactoring opportunities in the source code, in order to recommend \textit{pure} and \textit{root-canal} refactorings, \textit{i.e.,} behavior preserving code changes for the purpose of improving software quality. Figure~\ref{fig:motivation} depicts how our classification clusters the existing refactoring taxonomy reported in the literature \cite{kim2014empirical,Silva:2016:WWR:2950290.2950305,moser2007case,Tsantalis:2013:MES:2555523.2555539,palomba2017exploratory,vassallo2019large,pantiuchina2018developers,moser2006does}. As can be seen, our classification covers the majority of categories.

Also, it is important to note that existing studies, along with our manual analysis, have pointed out that refactoring can also be interleaved with other development tasks, such as updating functionalities, bug fixes, etc. 
We do not consider these categories (\textit{e.g.,} Bug Fix, Functional etc.) as part of our classification, since it can be performed using previous studies \cite{Hindle:2008:LCT:1370750.1370773,Hindle:2011:ATN:1985441.1985466,gharbi2019classification,Levin:2017:BAC:3127005.3127016}. More recently, Paixão et al. \cite{paixao2020behind} captured these additional refactoring categories (\textit{i.e.,} Bug Fix and Feature). In the future, we plan to extend our work to capture this taxonomy as well.

It is worth noting that there are many studies analyzing the impact of refactoring on 1) code smells, 2) internal quality attributes, and 3) external quality attributes, but our work focuses on the developer’s documentation, and not on the refactoring operations themselves and their impact. Our aim is to classify the intent. For example, when we classify a message stating the removal of duplicate code as a code smell, we are classifying purely the developer’s intent of removing duplicate code, so we are not claiming that the performed refactoring operations had an impact on only the removal of the code smell. In fact, these refactoring operations may also have an impact on other internal quality attributes, but such analysis is not what we are trying to achieve in this paper. Simply, we are classifying the developer’s intent and not the impact of the refactoring operations. The impact of refactoring operations has been heavily studied in literature and our study complements this effort by exposing what developers do care about when they refactor.

Finally, according to a recently published survey \cite{lacerda2020code}, refactoring is typically driven by intents that belong to the categories that we have used in this study.

\subsection{Benefits}

Commit messages are essential for not only comprehending code changes, but for many other aspects of software development, such as as classification of maintenance effort \cite{Hindle:2008:LCT:1370750.1370773, gharbi2019classification}, code change summarization \cite{mcburney2017towards}, files change-proneness and bug-proneness \cite{xia2016collective}, etc. For instance, recent studies have shown the feasibility of extracting insights of software quality from developers inline documentation. For instance, mining developers comments has unveiled how developers knowingly commit code that is either incomplete, temporary, or faulty. Such phenomenon is known as \textit{Self-Admitted Technical Debt} (SATD) \cite{potdar2014exploratory}. Similarly, our previous study has introduced \textit{Self-Affirmed Refactoring} (SAR) \cite{alomar2019can}, defined as developers explicit documentation of refactoring operations intentionally introduced as code change. Per analogy to SATD, SAR manifests as a positive phenomenon,  known to be one of the primary concepts to manage technical debt \cite{brown2010managing}. So, it  is  of  particular  interest  to  understand how the developer's intent to refactoring code leads to an adequate corrective action, \textit{i.e.}, SATD resolution, especially that recent studies focus on understanding how SATD is being removed \cite{7961515,7832911,maldonado2017empirical,zampetti2020automatically}.

When it comes to refactoring documentation, revealing the intents that are frequently pushing developers to refactor, is of a major importance for the community, especially that recent surveys have shown that refactoring tools are under-used, and developers are still manually refactoring their code \cite{kim2014empirical,murphy2012we}. And so, these patterns can narrow the scope of refactoring towards what developers consider to be relevant, in order to bridge the gap between refactoring tools and their adoption in practice. However, the identification of these SAR patterns, is human-intensive, subjective, and error-prone. Coping with the burden of manual analysis is the main goal of this paper, by initially detecting then classifying these SAR patterns, into the above-mentioned categories. Furthermore, the automated identification of these SAR patterns, is not straightforward, as these keywords, are not necessarily exclusive to refactoring. Even \textit{refactoring}, being the most intuitive keyword used to describe this activity, has been also found to be used out of its context  \cite{zhangpreliminary18}.   

Refactoring, just like any code change, has to be reviewed, before being merged into the code base. However, little is known about how developers \textit{perceive}  and \textit{review} refactoring during the code review process, especially that refactoring, by definition, is not intended to alter to the system's behavior, but to improve its structure, so its review may differ from other code changes. Yet, there is not much research investigating the proper documentation of refactoring, which can facilitate the process of reviewing it. Through the identification of these SAR patterns, many example of documented refactorings can be provided for future investigations and analysis. 


%% file: Tables/Pattern.tex
\begin{table*}[tbp]
\begin{center}
\caption{List of Self-Affirmed Refactoring (SAR) Patterns.}
\label{Table:Patterns}

\begin{adjustbox}{width=0.85\textwidth,center}
\begin{tabular}{llll}

\toprule
\textbf{Patterns} \\
         \midrule 
         (1) Refactor* & (30) Removed poor coding practice & (59) Change design  \\
         (2) Mov* & (31) Improve naming consistency & (60) Modularize the code  \\
         (3) Split* & (32) Removing unused classes & (61) Code cosmetics  \\
         (4) Fix* & (33) Pull some code up & (62) Moved more code out of  \\
         (5) Introduc* & (34) Use better name & (63) Remove dependency \\
         (6) Decompos* & (35) Replace it with  & (64) Enhanced code beauty \\
         (7) Reorganiz* & (36) Make maintenance easier & (65) Simplify internal design \\
         (8) Extract* & (37) Code cleanup & (66) Change package structure  \\
         (9) Merg* & (38) Minor Simplification  & (67) Use a safer method   \\
         (10) Renam* & (39) Reorganize project structures & (68) Code improvements \\
         (11) Chang* & (40) Code maintenance for refactoring & (69) Minor enhancement \\
         (12) Restructur* & (41) Remove redundant code & (70) Get rid of unused code \\
         (13) Reformat* & (42) Moved and gave clearer names to & (71) Fixing naming convention \\
         (14) Extend* & (43) Refactor bad designed code & (72) Fix module structure \\
         (15) Remov* & (44) Getting code out of & (73) Code optimization  \\
         (16) Replac* & (45) Deleting a lot of old stuff & (74) Fix a design flaw \\
         (17) Rewrit* & (46) Code revision & (75) Nonfunctional code cleanup  \\
         (18) Simplif* & (47) Fix technical debt & (76) Improve code quality\\
         (19) Creat* & (48) Fix quality issue & (77) Fix code smell \\
         (20) Improv* & (49) Antipattern bad for performances & (78) Use less code \\
         (21) Add* & (50) Major/Minor structural changes & (79) Avoid future confusion  \\
         (22) Modif* & (51) Clean up unnecessary code & (80) More easily extended  \\
         (23) Enhanc* & (52) Code reformatting \& reordering & (81) Polishing code \\
         (24) Rework* & (53) Nicer code / formatted / structure & (82) Move unused file away \\
         (25) Inlin* & (54) Simplify code redundancies & (83) Many cosmetic changes  \\
         (26) Redesign* & (55) Added more checks for quality factors & (84) Inlined unnecessary classes  \\
         (27) Cleanup & (56) Naming improvements & (85) Code cleansing \\
         (28) Reduc* & (57) Renamed for consistency & (86) Fix quality flaws \\
         (29) Encapsulat* & (58) Refactoring towards nicer name analysis & (87) Simplify the code \\
        \bottomrule
\end{tabular} 
\end{adjustbox}
\end{center}
\end{table*}

%% file: Tables/QACodeSmell.tex
\begin{table}[tbp]
\begin{center}
\caption{Quality Issues (Quality Attribute(s) \& Code Smell(s)) Extracted from SAR Commits.}
\label{Table:QA & Code smell}

\begin{adjustbox}{width=0.6\textwidth,center}
\begin{tabular}{lll}

\toprule
\textbf{Internal QA} &
\textbf{External QA} &
\textbf{Code Smell} \\
         \midrule 
    
    Inheritance (31.04\%) & Functionality (34.03\%) & Duplicate Code (43.52\%) \\ 
    Abstraction (30.63\%) & Performance (31.37\%) & Dead Code (24.84\%)  \\ 
    Complexity (14.30\%) & Compatibility (13.61\%) & Data Class (22.93\%) \\ 
    Composition (12.53\%)  & Readability (3.60\%) & Long Method (3.82\%) \\ 
    Coupling (3.81\%) & Stability (2.64\%) & Switch Statement (3.18\%) \\ 
    Encapsulation (3.61\%) & Usability (1.60\%) & Lazy Class (0.42\%) \\
    Design Size (2.11\%)  & Flexibility (1.58\%) & Too Many Parameters (0.42\%) \\
    Polymorphism (1.50\%) & Extensibility (1.54\%)  & Primitive Obsession (0.21\%) \\
    Cohesion (0.48\%)  & Efficiency (1.51\%) & Feature Envy (0.21\%) \\ 
    & Accuracy (1.05\%) & Blob Class (0.21\%) \\ 
    & Accessibility (1.04\%) & Blob Operation (0.21\%) \\
    & Robustness (0.78\%) &     \\ 
    & Testability (0.75\%) & \\ 
    & Correctness (0.65\%)& \\
    & Scalability (0.62\%) & \\
    & Configurability (0.56\%) & \\
    & Simplicity (0.55\%) & \\
    & Reusability (0.45\%) & \\
    & Reliability (0.43\%)& \\
    & Modularity (0.37\%) & \\
    & Maintainability (0.26\%) & \\
    & Traceability (0.26\%) & \\
    & Interoperability (0.24\%)  & \\
    & Fault-tolerance (0.16\%) & \\
    & Repeatability (0.07\%) & \\
    & Understandability (0.06\%) & \\
    &  Effectiveness (0.06\%) & \\
    & Productivity (0.06\%)& \\
    & Modifiability (0.03\%) & \\
    & Reproducibility (0.03\%) & \\
    & Adaptability (0.03\%) & \\
    & Manageability (0.01\%) & \\

        \bottomrule
\end{tabular} 
\end{adjustbox}

\end{center}
\end{table}

%% file: Sections/RelatedWork.tex
\section{Related Work}
\label{sec:RelatedWork}

In this section, we report studies related to developer's perception of refactoring and its documentation, along with the current state-of-the-art studies related to commit messages classification.

\subsection{Refactoring and its documentation}

A number of studies have focused recently on the identification and detection of refactoring activities during the software life-cycle. One of the common approaches to identify refactoring activities is to analyze the commit messages in versioned repositories. Stroggylos \& Spinellis \cite{stroggylos2007refactoring} searched words stemming from the verb \textit{\say{refactor}} such as \say{refactoring} or \say{refactored} to identify refactoring-related commits. Ratzinger et al. \cite{Ratzinger:2008:RRS:1370750.1370759,citeulike:2881658} also used a similar keyword-based approach to detect refactoring activity between a pair of program versions to identify whether a transformation contains refactoring. The authors identified refactorings based on a set of keywords detected in commit messages, and focusing on the following 13 terms in their search approach: \textit{refactor, restruct, clean, not used, unused, reformat, import, remove, replace, split, reorg, rename, and move}. 

Later, Murphy-Hill et al. \cite{murphy2012we} replicated Ratzinger's experiment in two open source systems using Ratzinger's 13 keywords. They conclude that commit messages in version histories are unreliable indicators of refactoring activities. This is due to the fact that developers do not consistently document refactoring activities in the commit messages. In another study, Soares et al. \cite{soares2013comparing} compared and evaluated three approaches, namely,  manual analysis, commit message (Ratzinger et al.'s approach \cite{Ratzinger:2008:RRS:1370750.1370759,citeulike:2881658}), and dynamic analysis (SafeRefactor approach \cite{Soares2009safetytool}) to analyze refactorings in open source repositories, in terms of behavioral preservation. The authors found, in their experiment, that manual analysis achieves the best results in this comparative study and is considered as the most reliable approach in detecting behavior-preserving transformations. 

In another study, Kim et al. \cite{kim2014empirical} surveyed 328 professional software engineers at Microsoft to investigate when and how they do refactoring. They first identified refactoring branches and then asked developers about the keywords that are usually used to mark refactoring events in commit messages. When surveyed, the developers mentioned several keywords to mark refactoring activities. Kim et al. matched the top ten refactoring-related keywords identified from the survey (\textit{refactor, clean-up, rewrite, restructure, redesign, move, extract, improve, split, reorganize, rename}) against the commit messages to identify refactoring commits from version histories. Using this approach, they found 94.29\% of commits do not have any of the keywords, and only 5.76\% of commits included refactoring-related keywords.

Peruma et al. \cite{peruma2019contextualizing} investigated how method, class, and package identifier names evolve and how this evolution was documented in the commit corresponding messages. By also analyzing their surrounding refactoring operations, the authors observed how names are influenced by their neighboring changes. They extended their work by considering the situation where a rename is applied to an identifier whose data type is changed \cite{peruma2020contextualizing}. Such case is of an interest to the authors as it indicates a potential change in the behavior its associated identifier.

Prior work \cite{Peruma:2018:EIW:3242163.3242169,liang2018,alomar2019can} has explored how developers document their refactoring activities in commit messages; this activity is called Self-Admitted Refactoring or Self-Affirmed Refactoring (SAR). In particular, 
 SAR indicates developers' explicit documentation of refactoring operations intentionally introduced during a code change.

The existence of such patterns unlocks more studies that question the developer's perception of quality attributes (\textit{e.g.,} coupling, complexity), typically used in recommending refactoring. For instance, AlOmar et al. \cite{alomar2019empiricalemse} identified which quality models are more in-line with the developer's vision of quality optimization when they explicitly mention in the commit messages that they refactor to improve these quality attributes. This study shows that, although there is a variety of structural metrics can represent internal quality attributes, not all of them can measure what developers consider to be an improvement in their source code. Based on their empirical investigation, for metrics that are associated with quality attributes, there are different degrees of improvement and degradation of software quality. 

\input{Tables/CommitClassification.tex}

\subsection{Commit Classification}
A wide variety of approaches to categorizing commits have been presented in the literature. The approaches vary between performing manual classification \cite{Silva:2016:WWR:2950290.2950305,cedrim2017understanding,Tsantalis:2013:MES:2555523.2555539,4686322,7180125,Chavez:2017:RAI:3131151.3131171}, to developing an automatic classifier \cite{mockus2000identifying,Hassan:2008:ACC:1363686.1363876,Mauczka2012}, to using machine learning techniques \cite{Hindle:2011:ATN:1985441.1985466,Levin:2017:BAC:3127005.3127016,article,5090025,5561540,levin2019towards,honel2019importance,McMillan:2011:CSA:2117694.2119646}  and developing discriminative topic modeling \cite{YAN2016296} to classify software changes. We summarize these state-of-the-art approaches in Table~\ref{Table:Commits Classification}.

Hattori and Lanza \cite{4686322} developed a lightweight method to manually classify history logs based on the first keyword retrieved to match four major development and maintenance activities: Forward engineering, Re-engineering,  Corrective engineering, and Management activities. Also, Mauczka et al. \cite{7180125} have addressed the multi-category changes manually using three classification schemes from existing literature. Tsantalis et al. \cite{Tsantalis:2013:MES:2555523.2555539} conducted a multidimensional empirical study on refactorings and  performed a systematic labeling of the commit messages to better understand the purpose of the applied refactorings. Silva et al. \cite{Silva:2016:WWR:2950290.2950305} applied a thematic analysis process to reveal the actual motivation behind refactoring instances after collecting all developers' responses. Further, a few studies \cite{cedrim2017understanding,Chavez:2017:RAI:3131151.3131171} propose the classification of refactoring instances as root-canal or floss refactoring through the use of  manual inspection. Yan et al. \cite{YAN2016296} used discriminative topic modeling techniques to automatically classifying software changes. 

Mockus \& Votta \cite{mockus2000identifying} designed an automatic classification algorithm to classify maintenance activities based on a textual description of changes. Another automatic classifier is proposed by Hassan \cite{Hassan:2008:ACC:1363686.1363876} to classify commit messages as a bug fix, introduction of a feature, or a general maintenance change. Mauczka et al. \cite{Mauczka2012} developed an Eclipse plug-in named Subcat to classify the change messages into Swanson's original category set (\textit{i.e.,} Corrective, Adaptive and Perfective \cite{Swanson:1976:DM:800253.807723}), with an additional category \say{Blacklist}. He automatically assessed if a change to the software was due to a bug fix or refactoring based on a set of keywords in the change messages. Hindle et al. \cite{Hindle:2008:LCT:1370750.1370773} performed a manual classification of large commits to understand the rationale behind these commits. Later, Hindle et al. \cite{5090025} proposed an automated technique to classify commits into maintenance categories using seven machine learning techniques. To define their classification schema, they extended Swanson's categorization \cite{Swanson:1976:DM:800253.807723} with two additional changes: Feature Addition, and Non-Functional. They observed that no single classifier is the best. Another experiment that classifies history logs was conducted by Hindle et al. \cite{Hindle:2011:ATN:1985441.1985466}, in which their classification
of commits involves the non-functional requirements (NFRs) a commit addresses. Since the commit may possibly be assigned to multiple NFRs, they used three different learners for this purpose along with using several single-class machine learners. Amor et al. \cite{article} had a
similar idea to \cite{5090025} and extended the Swanson categorization
hierarchically. They, however, selected one classifier (\textit{i.e.,} Naive Bayes) for their classification of code transactions. Moreover, maintenance requests have been classified using two different machine learning techniques (\textit{i.e.,} Naive Bayesian and Decision Tree) in \cite{5561540}. McMillan et al.  \cite{McMillan:2011:CSA:2117694.2119646} explored three popular learners to categorize software application for maintenance. Their results show that SVM is the best performing machine learner for categorization over the others. 

Levin and Yehudai  \cite{Levin:2017:BAC:3127005.3127016} automatically classified commits into three main maintenance activities using three classification models namely, J48, Gradient Boosting Machine (GBM), and Random Forest (RF). They found that the RF model outperforms the two other models (accuracy: 76\% versus 70\% and 72\%). In their extended work \cite{levin2019towards}, the RF model showed a promising accuracy of 76\%. More recently, a replicated study \cite{honel2019importance} of \cite{Levin:2017:BAC:3127005.3127016} introduced code density of a commit to study the purpose of a change. Using code-density based classification, they achieved up to 89\% accuracy for cross project commit classification using LogitBoost classifier. 

In this paper, we build on top of these techniques to leverage an automated identification and classification of SARs. Although the manual summarization of SAR is useful, it is considered as a time-consuming task because of the required manual effort to derive the list of patterns. Although much work has been done on automatically classifying commits in general, there is no currently automatic way to identify SAR patterns specifically. Several studies \cite{Hindle:2008:LCT:1370750.1370773,Levin:2017:BAC:3127005.3127016,article,4686322,5090025,Mauczka2012,7180125,YAN2016296,levin2019towards,honel2019importance} have discussed how to automatically classify change messages into Swanson’s general maintenance categories (\textit{i.e.,} Corrective, Adaptive, Perfective). Refactoring, in general, has been classified as a sub-type of \say{Perfective} in this maintenance category. Currently, there is no study that reports specific subcategories of refactoring extracted from real-world scenarios of commit messages and performs an automated classification of SAR commits. Therefore, in this paper, we push the Self-Affirmed Refactoring research a step forward by introducing an automatic classification approach to (1) determine whether a commit contains SAR or not (\textit{cf.}, Table~\ref{Table:Patterns}), and (2) classify SAR into its three categories (see Table~\ref{Table:QA & Code smell}). Compared with the pattern-based approach, our automated approach can identify more SAR patterns that can complement and extend the list of patterns identified in \cite{alomar2019can}. 

Further, in this work, we are detecting the indicators of refactoring to understand how developers document refactoring. We are not labelling refactoring operations themselves; we are instead labelling the commit messages that are found to contain refactoring operations. The existence of refactoring operations, in the studied commits, can be verified by running state-of-the-art tools, such as Refactoring Miner \cite{tsantalis2018accurate} and RefDiff \cite{silva2017refdiff} tools. Both of these studies indicated that their tool achieves high accuracy (precision of 98\% and 100\%, and recall of 87\% and 88\%, respectively), which gives us confidence to use one of these tools as a form of validation that the commits contain refactoring.

As can be seen in table~\ref{Table:Commits Classification}, commit messages are extensively used in existing literature to classify several maintenance-related tasks. Studies that focus on classifying bug and feature requests have used commit messages as a primary  source of information to generate high accuracy and applicable results. However, our approach is not restricted to a specific source of textual information. Future work could replicate our approach with other types of metadata, \textit{e.g.,} issue descriptions.

%% file: Tables/CommitClassification.tex
\begin{table*}
  \centering
	 \caption{Characteristics of Commit Classification Studies.}
	 \label{Table:Commits Classification}
\begin{sideways}
\begin{adjustbox}{width=\textheight,totalheight=\textwidth,keepaspectratio}
\begin{tabular}{lcllllll}\hline
\toprule
\bfseries Study & \bfseries Year  & \bfseries Manual/Automatic & \bfseries Classification Method &  \bfseries Category & \bfseries Machine Learning & \bfseries Training Size & \bfseries Result \\
\midrule
Mockus \& Votta \cite{mockus2000identifying} & 2000 & No/Yes & Automated Classifier & Maintenance Activities & N/A & 40 maintenance requests (8 participants) & Accuracy:  $\sim$ 61\% \\  \hline
Amor et al. \cite{article} & 2006 &  No/Yes  & Machine Learning & Swanson's category & NaiveBayes & 400 commits (1 participant) & Accuracy: 70\% \\  
& & & & Administrative & \\ \hline 
Hattori \& Lanza \cite{4686322} & 2008 &  No/Yes & Keywords-based Search & Forward Engineering & N/A & 1088 commits &  F-measure: 76\%  \\ 
& & & & Reengineering \\ 
& & & & Corrective Engineering \\ 
& & & & Management \\ \hline
Hassan \cite{Hassan:2008:ACC:1363686.1363876} & 2008 &  No/Yes & Automated Classifier & Bug Fixing & N/A & 18 commits (6 participants) & Agreement: 70\%   \\
& & & & General Maintenance \\
& & & & Feature Introduction \\
\hline
Hindle et al. \cite{Hindle:2008:LCT:1370750.1370773} & 2008 & Yes/No & Systematic Labeling & Swanson's category & N/A & 2000 commits & Not mentioned  \\ 
& & & & Feature Addition \\
& & & & Non-Functional \\ \hline 
Hindle et al. \cite{5090025} & 2009 &  No/Yes & Machine Learning & Swanson's category & J48 / NaiveBayes / SMO  & 2000 commits & F-measure: 51\%  \\ 
& & & & Feature Addition &  KStar / IBk / JRip / ZeroR & & Accuracy: 52\% \\ 
& & & & Non-Functional \\ \hline
Mahmoodian et al. \cite{5561540} & 2010 &  No/Yes & Machine Learning & Corrective \& Adaptive & NaiveBayes / ADtree & 1700 requests & Accuracy: 78\% \\ \hline 
Hindle et al. \cite{Hindle:2011:ATN:1985441.1985466} & 2011 &  No/Yes & Machine Learning & Non-Functional &  rule / decision trees / vector space & Not Mentioned & Receiver Operating  \\ 
&  & & & & SVM / CLR / HOMER / BR 
&  & Characteristic up to 80\% 
\\ \hline
Mauczka et al. \cite{Mauczka2012} & 2012 &  No/Yes  & Automated Classifier & Swanson's category & N/A & 21 commits
(5 participants) & Precision: 92\%  \\                      &  & & (Subcat tool) & Blacklist & & & Recall: 85\%  \\  \hline
Tsantalis et al. \cite{Tsantalis:2013:MES:2555523.2555539} & 2013 &  Yes/No & Systematic Labeling & Code Smell Resolution & N/A & Not Mentioned & Manual   \\ 
& & & & Extension \\
& & & & Backward Compatability \\
& & & & Abstraction Level Refinement \\ \hline
Mauczka et al. \cite{7180125} & 2015 &  Yes/No & Systematic Labeling & Swanson's category & N/A & 967 commits & Manual \\ 
& & & & Hattori \& Lanza category \cite{4686322} \\
& & & & Non-Functional \\ \hline
Yan et al. \cite{YAN2016296} & 2016 &  No/Yes &  Topic Modeling & Swanson's category & N/A & 80 commits 
(5 participants) & F-measure: 76\% \\ \hline
Silva et al. \cite{Silva:2016:WWR:2950290.2950305} & 2016 & Yes/No & Systematic Labeling & Refactoring's Motivation & N/A & N/A & Manual \\ \hline
Chavez et al. \cite{Chavez:2017:RAI:3131151.3131171} & 2017 &  Yes/No & Systematic Labeling & Floss Refactoring & N/A & sample of 2119 & Manual \\ 
& & & & Root-canal Refactoring \\ \hline
Cedrim et al. \cite{cedrim2017understanding} & 2017 &  Yes/No & Systematic Labeling & Floss Refactoring & N/A & part of sample of 2584 & Manual \\ 
& & & & Root-canal Refactoring  \\ \hline
Levin \& Yehudai \cite{Levin:2017:BAC:3127005.3127016} & 2017 & No/Yes & Machine Learning & Swanson's category &  J48 / GBM / RF & 1151 commits & Accuracy: 76\% \\
\hline 
Levin \& Yehudai  \cite{levin2019towards} & 2019 & No/Yes & Machine Learning & Swanson's category & J48 / GBM / RF & 1151 commits & Accuracy: 76\% 
\\ \hline
Honel et al. \cite{honel2019importance} & 2019 & No/Yes & Macine Learning & Swanson's category & LssvmRadical / SVM /  GBM     & 1151 commits & Accuracy: up to 89\% \\ 
& & & & & xgbTree / LDA / MDA / NN / avNNet  & & \\ 
& & & & & C5.0 / RF / Naive Bayes / LogitBoost & & \\\hline
\textbf{This work} & & \textbf{No/Yes} & \textbf{Machine Learning} & \textbf{SAR \& non-SAR} &  \textbf{LR / RF / GBM / DJ / SVM} &  \textbf{1823 commits} (two-class) & \textbf{Accuracy: 98\%} \\
& & & & \textbf{Internal QA}   &  \textbf{LD-SVM / NN / APM / BPM} & &  \textbf{F-mesaure: 98\%} \\
& & & & \textbf{External QA} &  &  \textbf{1044 commits} (multiclass) &  \textbf{Accuracy: 93\%}  \\
& & & & \textbf{Code Smell} & & & \textbf{F-mesaure: 93\%} \\

\bottomrule
\end{tabular}
\end{adjustbox}
\end{sideways}
\end{table*}

%% file: Sections/Approach.tex
\section{Approach}
\label{sec:Approach}

\begin{figure*}[htbp]
\centering 
\includegraphics[width=1\textwidth]{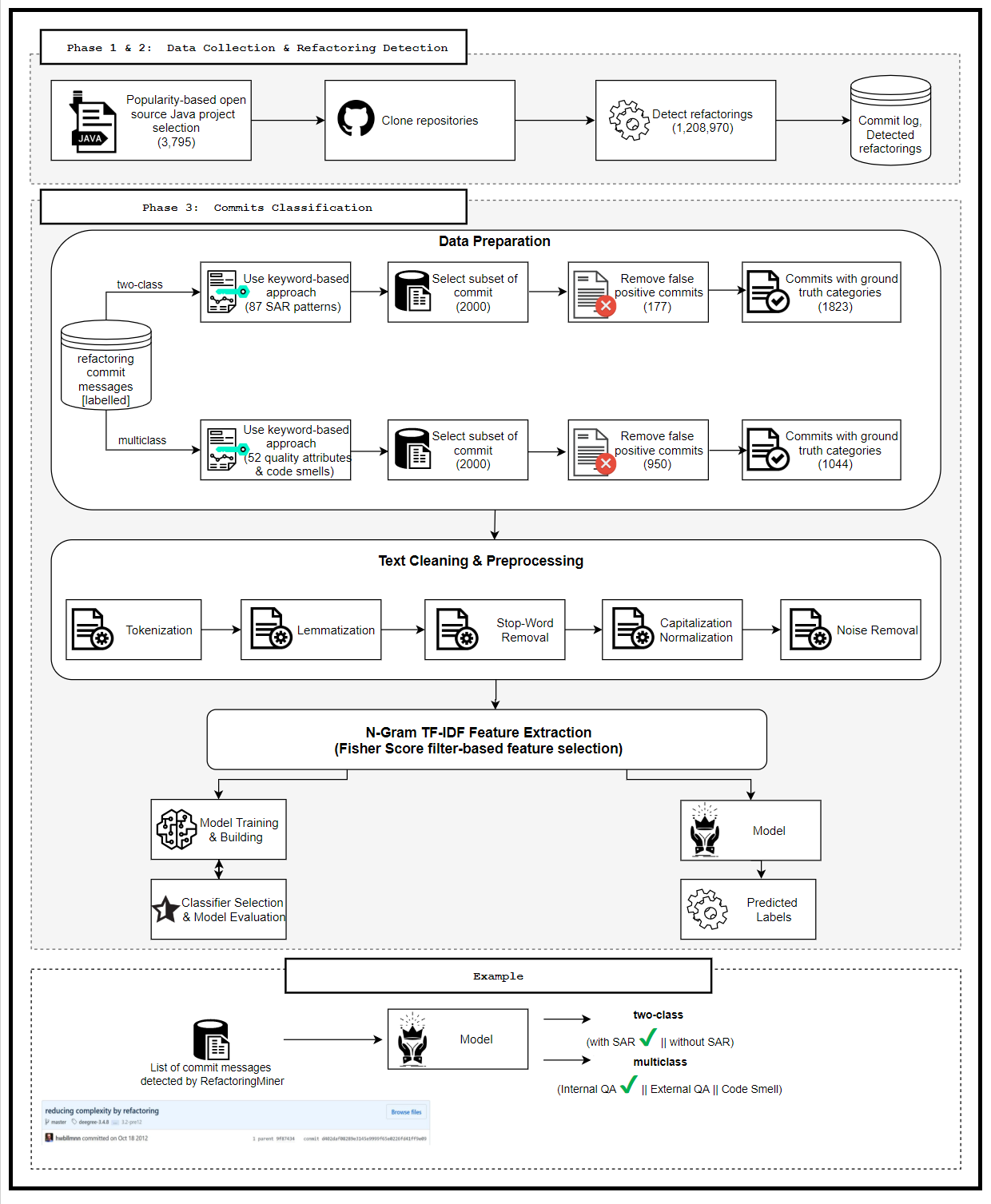}
\caption{Overall Classification Framework.} 
\label{fig:approach_overview}
\end{figure*}
\begin{figure*}[htbp]
\centering 
\includegraphics[width=1\textwidth]{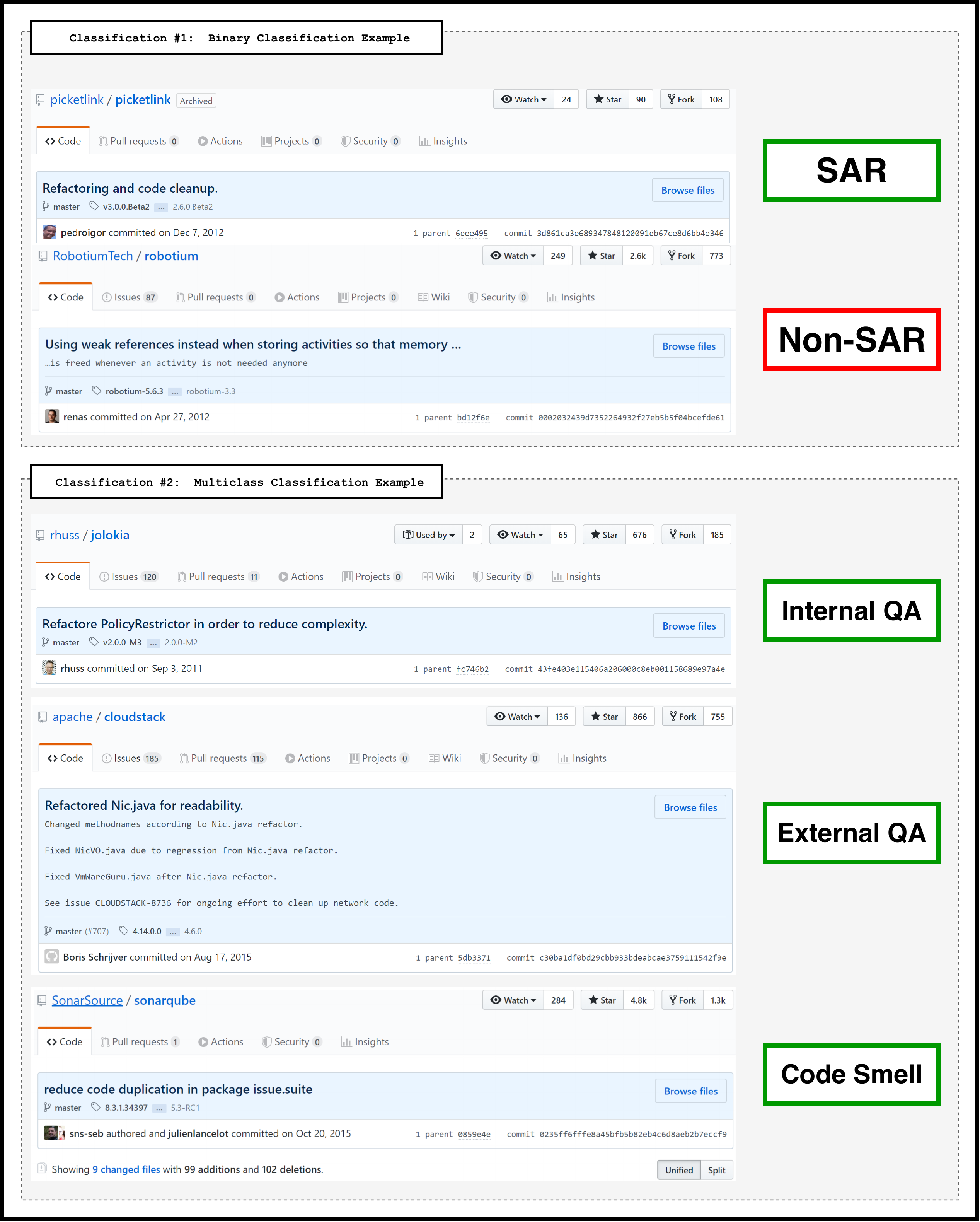}
\caption{Commit Message Examples for Binary and Multiclass Classification.} 
\label{fig:example}
\end{figure*}
In this section, we first provide an overview of our approach. Then, we elaborate on the technical details of the adopted classification technique, in the following subsection. The overview of our approach is depicted in Figure~\ref{fig:approach_overview}, and a sample of commit messages is demonstrated in Figure~\ref{fig:example}.

\subsection{Data Collection}
To collect the necessary commits, we refer to an existing large dataset of links to GitHub repositories \cite{allamanis2013mining}. We perform an initial filtering, using Reaper \cite{munaiah2017curating}, to only navigate through well-engineered projects while verifying that they were Java-based; the only language supported by Refactoring Miner. The authors of this dataset classified \say{well-engineered software projects} based on the projects' use of software engineering practices such as documentation, testing, and project management. So, we ended up reducing the number of selected projects from 57,447 to 3,795.

Using \say{well-engineered} and \say{well-documented} kind of interchangeably - although we acknowledge the potential value of having a more diverse set of projects, and our findings may not extend to projects that are not as well-documented, because our primary research methods rely on documentation, we chose to focus on projects that would be likely to have high-quality documentation (\textit{i.e.,} commit messages) consistently available.

\subsection{Refactoring Detection}
To extract the entire refactoring history in each project, we use the popular refactoring mining tool, \textit{i.e.,}  Refactoring Miner \cite{Silva:2016:WWR:2950290.2950305}. 
Our choice to use Refactoring Miner is justified by the fact that it achieved the highest accuracy in detecting refactorings compared to the state-of-the-art available tools, with a precision of 98\% and recall of 87\% \cite{Silva:2016:WWR:2950290.2950305,tsantalis2018accurate} along with being suitable for our study that requires a high degree of automation in data mining. In this phase, We collect a total of 1,208,970 refactoring operations from 322,479 commits, applied during a period of 23 years (1997-2019). 

\subsection{Overall Framework}
In a nutshell, the goal of our work is to automatically identify then classify commit messages containing refactoring documentation, \textit{i.e.}, Self-Affirmed Refactorings, (for the sake of simplicity, we refer to them as SAR). Our approach takes as input, a commit message, and makes a binary decision on whether it contains SAR or not. If a SAR is detected, it classifies it into one of of three common categories: (\textit{i}) internal quality attribute (\textit{ii}) external quality attribute, and (\textit{iii}) code smell 
\cite{alomar2019can}. The overall framework of our approach is depicted in Figure~\ref{fig:approach_overview}. We formulate a two-phased approach that consists of a model building phase and a prediction phase. In the model building phase, our goal is to build a model from a corpus real world documented refactoring operations (\textit{i.e.,} commit messages). In the prediction phase, the model created in the previous phase will be used to predict categories of new refactoring-related commit messages.

Our framework takes commit messages along with their ground truth categories obtained by manual inspection as input for the training procedure extracted from different projects, provided by a previous study \cite{alomar2019can}. Based on this input, the commit messages are preprocessed, allowing for informative featurization. Next, for each commit message, we extract features (\textit{i.e.,} words) to create a structured feature space. Then, we use the extracted features to build the training set. In total, we experimented 9 commonly used classifiers to evaluate our model for prediction. We selected these classifiers as they are typically used in previous commit classification studies as well as several software engineering classification/prediction problems \cite{Hindle:2011:ATN:1985441.1985466,Levin:2017:BAC:3127005.3127016,article,5090025,5561540,levin2019towards,honel2019importance}, 
 as outlined in Table \ref{Table:Commits Classification}. 
After training all models, we use a testing set to challenge the performance. Since the model has already learned the vocabulary of N-Gram (discussed in Section~\ref{sec:n-gram}) 
and their weights from the training dataset, we extract features from the test data based on that vocabulary and weights, and input them to the model. Finally, the classifier will output the predicted label for each tested commit message. 

\subsection{Commit Classification}
Our classification process has five main phases: (1) data preparation, (2) text cleaning and preprocessing, (3) feature extraction using N-Gram, (4) model training and building, and (5) classifier selection and model evaluation. Since a commit message is written in plain text, we follow the approach provided by Kowsari et al. \cite{kowsari2019text} that discussed a recent trend in text classification techniques and algorithms.  

\subsubsection{Data Preparation}


Our goal is to provide the classifier with sufficient commits that represent the categories analyzed in this study. Since the number of candidate commits to classify is large, we cannot manually process them all, and so we need to randomly sample a subset while making sure it equitably represents the featured classes, \textit{i.e.,} categories. Since an imbalanced training dataset or class starvation (\textit{i.e.,} not having adequate instances of a certain class) could worsen the performance of the model \cite{Levin:2017:BAC:3127005.3127016,levin2019towards}, we make sure that the classes for two-class (\textit{i.e.,} with or without SAR) and multiclass (\textit{i.e.,} Internal QA, External QA, and Code Smell) classification problems are equally distributed when preparing the data for the training (cf. Table~\ref{Table:Instances per class (train, test)}). The classification process has been performed by the authors of the paper. To approximate the needed number of commits to add, we reviewed the thresholds used in the studies related to commit classification (see Table~\ref{Table:Commits Classification}). The highest number of commits used in comparable studies was around 2,000 commits \cite{Hindle:2008:LCT:1370750.1370773,5090025,Chavez:2017:RAI:3131151.3131171}. Thus, we select a sample of 2,000 commits from 3,795 projects for each classification model.  
Below we detail the manual analysis of the data we use for our classification.


\input{Tables/TrainTest.tex}

For data preparation, building the ground truth is challenging since we are looking for a particular set of commits. To prune the search space, we started with using an existing dataset of commits \cite{alomar2019can}, manually inspected and validated for containing refactoring operations and an associated description at the commit message. We intend to build our own dataset by choosing a subset of this dataset, in a way to serve the purpose of the binary and multiclass classification. 

To prepare the dataset of the binary classification, we need to create two groups of commits, \textit{i.e.,} commits with or without SAR. The first group (with SAR) is created by randomly sampling commits, previously known to contain SAR patterns listed in Table~\ref{Table:Patterns}. We further perform another round of individual verification by the authors before adding them to the group. Commits for which there was no full agreement by the authors were excluded from our dataset. The second group (without SAR) can be easily created by randomly choosing commit messages that simply do not contain these SAR patterns, but since we do want to strengthen our decision boundary, we intend to choose commits that are closest to neighboring regions between the two classes. To do so, for each commit from the first group (with SAR), we locate the set of its contiguous commits (committed either before or after), and performed by the same committer, then we randomly sample one of them to be added to the second group (without SAR), after manually verifying that it does not contain any description of a refactoring activity. 

For the multiclass classification, we build it by making sure the chosen commits belong to one of the three categories listed in Table~\ref{Table:QA & Code smell}. To avoid involving our interpretation, it is important to note that the description of the categories listed in Table~\ref{Table:QA & Code smell} needs to be explicit in the commit messages. We used stratified sampling to select 2,000 commit messages for manual classification, divided equally for each stratum. To ensure that these commits reported developers’ intention to perform refactoring, and to improve quality attributes or fix code smells, we inspected these commits to remove false positives. 

To avoid having false positive commits, we applied the filtering to narrow down the commit messages eliminating the ones that are less likely to be classified as self-affirmed refactoring. We designed the filtering to help ensure that we only trained the algorithm on higher-quality commit messages \cite{jiang2017automatically}.

We followed the process from existing papers in filtering commit messages \cite{Mauczka2012,fu2015automated,da2017using}. For example, Fu et al. \cite{fu2015automated} filtered out short commit messages. Mauczka et al. \cite{Mauczka2012} used the \say{Blacklist} category to filter all commits, which underlying modifications were not carried out by humans or which do not actually include any source code modifications. In our work, we apply five filtering heuristics to narrow down the commit messages eliminating the ones that are less likely to be classified as SAR. It is important to note that we removed short commit messages from the training, but not from the testing set because (1) short commit messages do not contain enough information and do not clearly describe the purpose of code change , and (2) we want to train the classifier on well-documented commit messages, and label commits that contain enough information about refactorings. Prior study has pruned short commit messages since these will be noise for the classifiers, and they did not record the cause of the changes \cite{fu2015automated}. Some criteria we used for filtering were as follows: 

\begin{itemize}
    \item If a commit contains an alternative form of the word \say{refactor} such as \say{re-factor*}, the commit was  classified as SAR commit. 
    \item If a commit message contains a pattern that is in a slightly different form of one of the patterns, such as \say{simplify the code} and \say{simplify code}, the commit was classified as SAR commit. 
    \item Commits that were either too short or ambiguous were discarded. Some examples of hard-to-classify commit messages are: \say{\textit{Solr Indexer ready}}\footnote{https://github.com/01org/graphbuilder}, \say{\textit{allow multiple collections}}\footnote{https://github.com/0install/java-model}, and \say{\textit{Auto configuration of AgiScripts}}\footnote{https://github.com/1and1/attach-qar-maven-plugin}.
    \item If one commit could belong to more than one class, it was excluded.
    \item If the quality attribute is a part of the identifier name, the commits were excluded, \textit{e.g.,} \say{\textit{SONARJS-541 Precise issue location for ExpressionComplexity (S1067)}}.
    We discarded this commit because \say{complexity} is referring to a part of a class name and not a quality attribute. 
    
\end{itemize}

The above-mentioned examples of ambiguous commit messages prevent us from being confident, and hence, for each discarded commit message, we randomly sampled another replacement. We repeated this process until we found the commit message that we were able to confidently classify. 
Because of the random nature of the process, some classes were saturated faster than others, so we kept increasing the number of instances only for the underrepresented classes, until we find the right balance between all classes. 
The criteria listed above reduced the number of commits and helped us focus on the most insightful commit messages. For the binary classification, 177 commits were removed because of them either being short or ambiguous. Also, in our case, any message with less than 7 characters was too short for us to decide. The evaluation resulted in keeping 1,823 commits and 1,044 commits, respectively for two-class and multiclass classification problems. To mitigate the risk of having a biased dataset and to inspect the level of agreement of the manual classification, we extract stratified sample of our dataset that are classified by the first author, and have these sample commits independently classified again by the second author. Particularly,
similar to \cite{levin2019towards}, in order to inspect manual classification agreement, we randomly classified a 10\% sample of commits, \textit{i.e.,} 186 and 105 commits out of the 1823 and 1044 for two-class and multiclass classification problems, respectively. This quantity roughly equates to a sample size with a confidence level of 95\% and a confidence interval of 8.
We used Cohen's Kappa coefficient \cite{cohen1960coefficient} to evaluate the inter-rater agreement level for the categorical classes. We achieved an agreement level of 0.96 for the two-class classification, and 0.87 for multiclass classification. According to Fleiss et al. \cite{fleiss1981measurement}, these agreement values are considered to have an almost \textit{perfect agreement} (\textit{i.e.}, $0.81 – 1.00$).

The result of this classification is available in the reproduceability package of this work, thus, it can be reused and extended \cite{SAR2019WEB}. 

\subsubsection{Text Cleaning \& Preprocessing}
After the data preparation phase, we applied a similar methodology explained in \cite{kowsari2019text,kochhar2014automatic} for text pre-processing. In order for the commit messages to be classified into correct categories, they need to be preprocessed and cleaned; put into a format that the classification algorithms will process. This way, the noise will be removed, allowing for informative featurization. To extract features (\textit{i.e.,} words), we preprocess the text as follows: 

\begin{itemize}
    \item \textbf{Tokenization:} The goal of tokenization is to investigate the words in a sentence. The tokenization process breaks a stream of text into words, phrases, symbols, or other meaningful elements called tokens \cite{kowsari2019text}. In our work, we tokenize each commit by splitting the text into its constituent set of words. We also split tokens on special characters (\textit{e.g.,} the string \say{package-level} would be separated into two tokens, \say{package} and \say{level}).
    \item \textbf{Lemmatization:} The lemmatization process either replaces the suffix of a word with a different one or removes the suffix of a word to get the basic word form (lemma). In our work, the lemmatization process involves sentence separation, part-of-speech identification, and generating dictionary form. We split the commit messages into sentences, since input text could constitute a long chunk of text. The part-of-speech identification helps in filtering words used as features that aid in key-phrase extraction. Lastly, since the word could have multiple dictionary forms, only the most probable form is generated.  
    \item \textbf{Stop-Word Removal:} Stop words (\textit{i.e.,} words and common English words such as \say{is}, \say{are}, \say{if}, etc) are removed since they do not play any role as features for the classifier \cite{saif2014stopwords}.  
    \item \textbf{Capitalization Normalization:} Since text could have a diversity of capitalization to form a sentence and this could be problematic when classifying large commits, all the words in the commit messages are converted to lower case and all verb contractions are expanded.
    \item \textbf{Noise Removal:} Special characters and numbers are removed since they can deteriorate the classification. More specifically, we remove all numeric characters, unique and duplicate special characters, email addresses and URLs. 
\end{itemize}

\input{Tables/BinaryClassification.tex}

\input{Tables/MultiClassClassification.tex}

\subsubsection{Feature Extraction Using N-Gram} 
\label{sec:n-gram}
After cleaning and preprocessing the text, we apply feature extraction to extract only the most useful information from text strings to differentiate classes in both classification problems. In particular, we selected the N-Gram technique for feature extraction. The N-Gram technique is a set of \textit{n-word} that occurs in a text set and could be used as a feature to represent that text \cite{kowsari2019text}. In general, N-Gram term has more semantic than an isolated word. Some of the keywords (\textit{e.g.,} \say{improve}) do not provide much information when used on its own. However, when collecting N-Gram from commit message (\textit{e.g.,} \textit{Refactor:Remove redundant method names, extract method, improve usability}), the keyword \say{improve} clearly indicates that this is a SAR commit. In our classification, we use bigrams since it is very common to enhance the performance of text classification \cite{tan2002use}, and we select Fisher Score filter-based feature selection \cite{duda2012pattern,gu2012generalized} to \textit{featurize} text and manage the size of the text feature vector, similar to \cite{kochhar2014automatic}. As for the weighting function, we used the standard Term Frequency-Inverse Document Frequency (TF-IDF) \cite{manning2008schu} due to its popularity in the research community (the value for each N-Gram is its TF score multiplied by its IDF score). Thus, each preprocessed word in the commit message is assigned a value which is the weight of the word computed using this weighting scheme. TF-IDF gives greater weight (\textit{e.g.,} value) to words which occur frequently in fewer documents rather than words which occur frequently in many documents. 

\subsubsection{Model Training and Building}
In this phase, we performed the 10-fold cross-validation technique to assess the variability and reliability of the classifier. Specifically, for each of the classification methods, we combined the commit messages into a single large dataset. Then, we split the dataset into ten folds, where each fold contained an equal proportion of commit messages. Thereafter, we performed ten evaluation rounds with different testing dataset in which nine folds were used as training dataset and the remaining one of the ten folds is used as the testing dataset. We aggregated the results of the ten evaluation rounds and reported the average performance for each classifier.

\subsubsection{Classifier Selection and Model Evaluation}
Selecting the proper classifier for optimal classification of the commits is a rather challenging task \cite{fernandez2014we}. Best practices suggest that developers document their commits by providing a commit message along with every commit they make to the repository. These commit messages are usually written using natural language, and generally convey some information about the commit they represent. In this study, we are dealing with two-class and multiclass classification problems since the commit messages are categorized into two and three different types as explained in Table~\ref{Table:Patterns} and~\ref{Table:QA & Code smell}, respectively. Because we have a predefined set of categories, our approach relies on supervised machine learning algorithms to assign each commit message to one category.  
Since it is very important to come up with an optimal classifier that can provide satisfactory results, several studies have compared several classifiers such as K-Nearest Neighbor (KNN), Naive Bayes Multinomial, Gradient Boosting Machine (GBM), and Random Forest (RF) in the context of commit classification into similar categories \cite{Levin:2017:BAC:3127005.3127016, levin2019towards, kochhar2014automatic}. These studies found that Random Forest (RF) achieves high performance. 
We investigated each classifier ourselves using common statistical measures (\textit{precision, recall, accuracy, and F-measure}) of classification performance to compare each. It is important to note that the calculation of F-measure for multiclass classification is not supported by Azure Machine Learning (Azure ML). Thus, to facilitate comparison and to have all statistical measures that are consistent with two-class classification, we compute F-measure for multiclass in terms of precision and recall  using the following formula: 

\begin{equation}
F = 2*\left(\frac{Precision * Recall}
       {Precision + Recall}\right)
\end{equation}
where Precision (P) and Recall (R) are calculated as follows:
\begin{equation*}
P= \frac{tp}{tp+fp} , R= \frac{tp}{tp+fn}
\end{equation*}
It is worth noting that a few models that we consider are inherently binary classifiers. In order to adjust for multiclass classification, each classifier applies the One-vs-All strategy for issues that require multiple output classes \cite{Lorena2009}. Thus, to ensure fairness, we use One-vs-All strategy for multiclass classification when using the following five classifiers: Gradient Boosted Machine (GMB) \cite{friedman2001greedy}, Support Vector Machine (SVM) \cite{wu2008top}, Locally Deep SVM (LD-SVM) \cite{jose2013local}, Averaged Perceptron Method (APM) \cite{collins2002discriminative}, and Bayes Point Machine (BPM) \cite{herbrich2001bayes}. The remaining classifiers, consider in this study, are: Logistic Regression (LR) \cite{andrew2007scalable}, Random Forest (RF) \cite{prinzie2008random}, Decision Jungle (DJ) \cite{shotton2013decision}, and Neural Network (NN) \cite{hansen1990neural}. Our experiment is conducted using Microsoft Azure Machine Learning (Azure ML \cite{mund2015microsoft}), as it provides a built-in web-service once the classification models are deployed.

%% file: Tables/TrainTest.tex
\begin{table}[ht]
\begin{center}
\caption{Number of Instances per Class.}
\label{Table:Instances per class (train, test)}
\begin{adjustbox}{width=.6\columnwidth,center}
\begin{tabular}{lccc}\hline
\toprule
\bfseries Dataset & \bfseries with SAR & \bfseries without SAR   \\
\midrule
1,823 instances & 912 & 911 \\
\midrule
\bfseries Dataset & \bfseries Internal QA & \bfseries External QA & \bfseries Code Smell \\ 
\midrule
1,044 instances & 348 & 348 & 348\\
\bottomrule
\end{tabular}
\end{adjustbox}
\end{center}
\end{table}

%% file: Tables/BinaryClassification.tex
\begin{table*}[h]
\begin{center}
\caption{Performance of Different Classifiers (Binary Classification).}
\label{Table:binary classifiers}
\begin{adjustbox}{width=.9\columnwidth,center}
\begin{tabular}{lcccc}\hline
\toprule
\bfseries Classifier & \bfseries Precision & \bfseries Recall & \bfseries Accuracy  & \bfseries F-measure   \\
\midrule
Logistic Regression & 0.98 & 0.93 & 0.96 & 0.95\\
\rowcolor{lightgray} Random Forest & 0.98 & 0.98 & 0.98 & 0.98 \\
\rowcolor{lightgray} Gradient Boosted Machine & 0.98 & 0.98 & 0.98 & 0.98  \\
 Decision Jungle & 0.97 & 0.94 & 0.95 & 0.95  \\
Support Vector Machine & 0.96 & 0.94 & 0.95 & 0.95 \\
Locally Deep SVM & 0.97 & 0.93 & 0.95 & 0.95 \\
Neural Network & 0.98 & 0.92 & 0.95 & 0.95 \\
Averaged Perceptron Method & 0.97 & 0.93 & 0.95 & 0.95 \\
Bayes Point Machine & 0.83 & 0.85 & 0.84 & 0.84 \\

\bottomrule
\end{tabular}
\end{adjustbox}
\end{center}

\end{table*}

%% file: Tables/MultiClassClassification.tex
\begin{table*}[h]
\begin{center}
\caption{Performance of Different Classifiers (Multiclass Classification).} 
\label{Table:multi class classifiers}
\begin{adjustbox}{width=.9\columnwidth,center}
\begin{tabular}{lcccc}\hline
\toprule
\bfseries Classifier & \bfseries Precision & \bfseries Recall & \bfseries Accuracy  & \bfseries F-measure   \\
\midrule
\rowcolor{lightgray} Logistic Regression & 0.93 & 0.93 & 0.93 & 0.93\\
\rowcolor{lightgray} Random Forest & 0.93 & 0.93  & 0.93 & 0.93  \\
\rowcolor{lightgray} One-vs-All Gradient Boosted Machine & 0.93 & 0.93 & 0.93 & 0.93 \\
Decision Jungle & 0.89 & 0.88 & 0.88 & 0.88 \\
One-vs-All Support Vector Machine & 0.91 & 0.91 & 0.91 & 0.91 \\
One-vs-All Locally Deep SVM & 0.90 & 0.90 & 0.90 & 0.90 \\
Neural Network & 0.91 & 0.91 & 0.91 & 0.91\\ 
One-vs-All Averaged Perceptron Method & 0.91 & 0.90 & 0.90 & 0.91 \\
One-vs-All Bayes Point Machine & 0.83 & 0.83 & 0.83 & 0.83 \\
\bottomrule
\end{tabular}
\end{adjustbox}
\end{center}
\end{table*}

%% file: Sections/Results.tex
\section{Results \& Discussion}
\label{sec:ResultsDiscussions}

In this section, we conduct an empirical study to assess the performance of our approach. To evaluate different commit classification models, we used standard statistical measures to measure the performance of the classification (\textit{Precision, Recall, Accuracy and F-measure}). In the following, we report the results of our research questions. 

\textbf{Replication package.} We provide our comprehensive experiments package available in \cite{SAR2019WEB} 
to further replicate and extend our study.

\subsection{\textbf{RQ1: Is it possible to accurately perform two-class and multiclass SAR classification using our machine learning technique?}}
As shown in previous work \cite{alomar2019can}, SAR can be extracted from commit messages. However, there is a lack of automatic techniques to classify them. In this work, we performed an automated approach to classify SAR to determine if the classification using machine learning techniques can result in high accuracy. A comparison between classification algorithms is reported in Table~\ref{Table:binary classifiers} and~\ref{Table:multi class classifiers}. The best performing model was used to classify the test dataset. Based on our findings, the F-measure of Random Forest (RF) and Gradient Boosting Machine (GBM) are respectively 98\% and 98\% which are clearly higher than their competitors for the two-class classification. 
For the multiclass, in addition to RF and GBM, Logistic Regression (LR) outperforms the other models with F-measure of 93\%. 
Figures~\ref{Chart:Visualization of the Precision for Different Classifiers},~\ref{Chart:Visualization of the Recall for Different Classifiers}, and~\ref{Chart:Visualization of the F1-measure for Different Classifiers} show the detailed performance for the best multiclass classifiers. 

Random Forest and boosting learning machines belong to the family of ensemble learning machines, and have typically yielded superior predictive performance mainly due to the fact that they both aggregate several learnings. As for Logistic Regression,  the fact that Logistic Regression achieves comparable performance as 
Random Forest and Boosting can be explained by the fact that the underlying true model for the text data has an inherent structure that matches the logistic regression assumption. 
 
\begin{figure*}[ht]
\centering 
\includegraphics[width=0.9\linewidth]{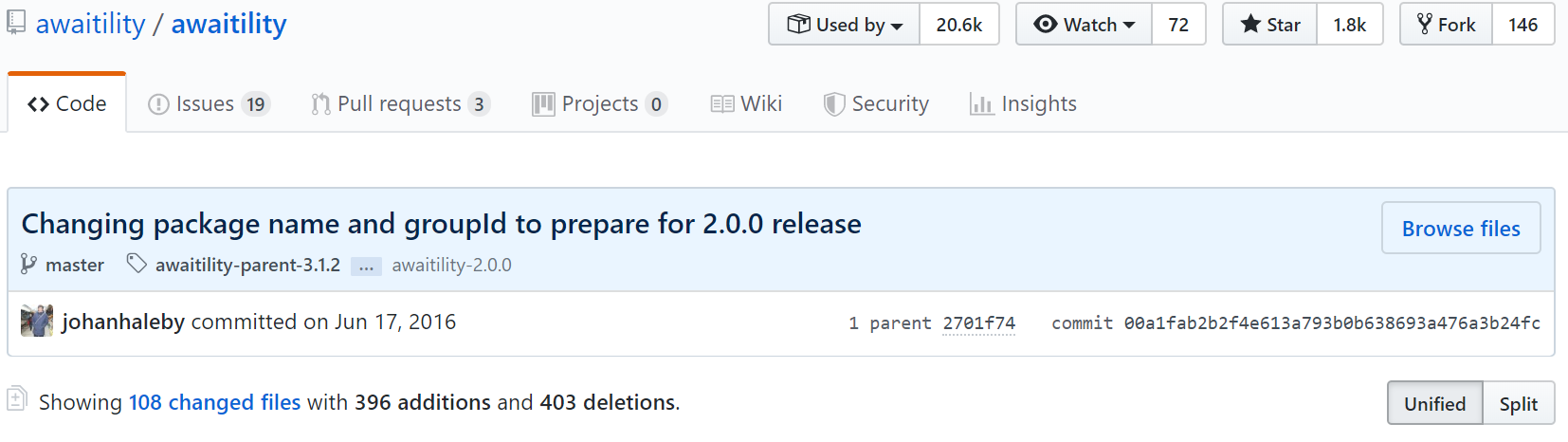}
\caption{Example of Refactoring Commit Message that Confused the Classifiers (Two-class).}
\label{fig:github-twoclass}
\vspace{0.75cm}
\centering 
\includegraphics[width=0.9\linewidth]{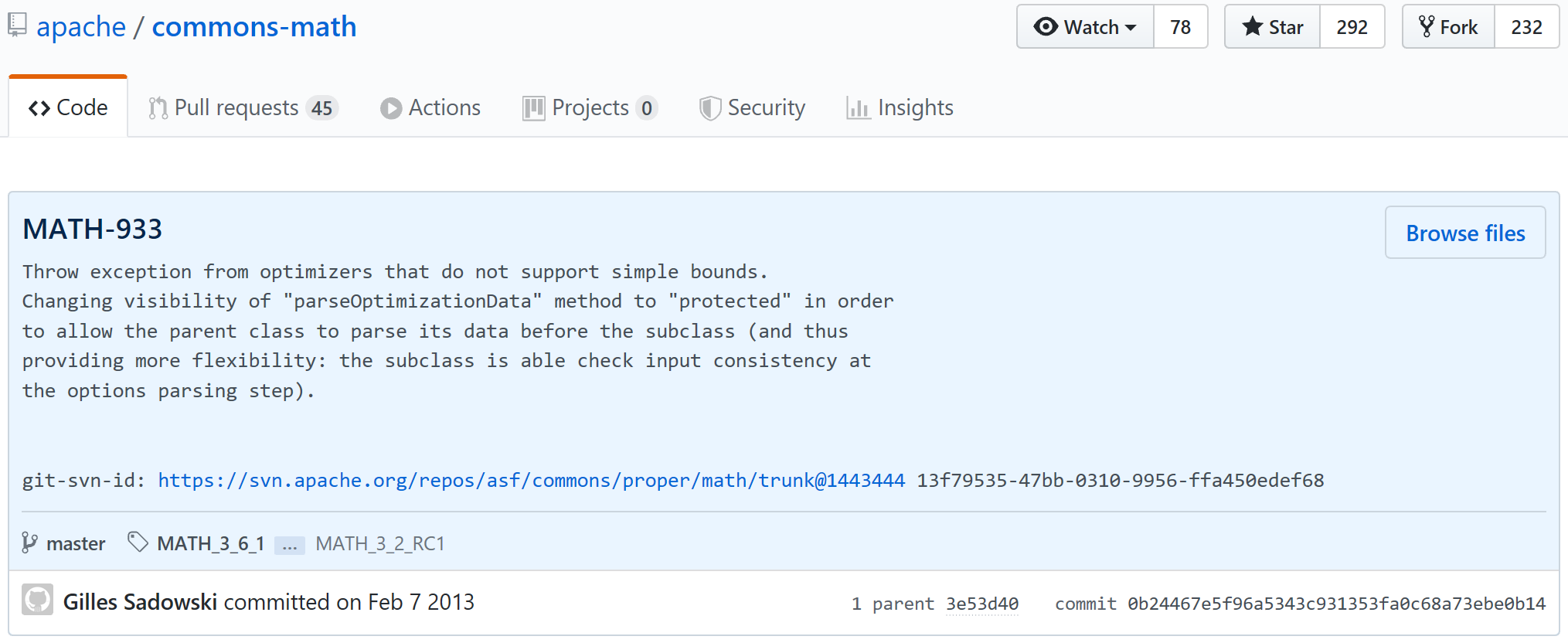}
\caption{Example of Refactoring Commit Message that Confused the Classifiers (Multiclass).}
\label{fig:github-multiclass}
\end{figure*}

Another observation with regard to the classifiers accuracy is that few of the classifiers we considered in our study (GBM, SVM, LD-SVM, APM, and BPM) are inherently binary classifiers, and we used the One-vs-All strategy to adapt them for multiclass. Hence, these classifiers give us higher accuracy when performing binary classification compared to multiclass classification (98\% vs 93\%, respectively).
 Another reason for getting a different accuracy value when identifying multiclass labels vs two-class is that some commit messages could potentially belong to multiple categories. Hence, the machine learning classifiers, considered in this study, got confused when classifying such commit messages. Figures~\ref{fig:github-twoclass} and~\ref{fig:github-multiclass} show two cases of commit messages that confused the classifiers when performing two-class and multiclass classifications, respectively. The first commit message (Figure \ref{fig:github-twoclass}) contains a pattern (\textit{i.e.,} changing package name) that is a synonym of the patterns \say{renam*} or \say{use better name}. The second commit message (Figure \ref{fig:github-multiclass}) contains more code element-related keywords such as \say{method} or \say{class} tend to be classified as \say{Code Smell} since code smell-related commits usually contain more description of the code elements that need to be optimized. This commit example targets to improve the flexibility at the design phase, which should be classified as \say{External QA}.  

Moreover, it is important to note that the classes used in this study categorize mainly the refactoring documentation and do not reflect the overall activities of the commits. Hence, these commit messages do not strictly contain refactoring code changes, especially that we noticed that refactoring tends to be interleaved with other software engineering tasks, such as fixing bugs, migrating type changes etc. Therefore, it is important to consider such \textit{context} to better understand the intention behind the application of refactoring, and this will be our main future research direction. 


\begin{tcolorbox}
\textit{Summary}. We find that our approach is accurately identifying the SAR patterns and the three common quality improvements with an F-measure of 98\% and 93\% for the  two-class and multiclass classification problems, respectively. 
\end{tcolorbox}

\input{Charts/DifferentApproachVisualization_V2.tex}

\subsection{\textbf{RQ2: How effective is our machine learning approach in classifying SAR?}}
The main goal of this research is to propose an automatic approach to classify SAR commits that can effectively outperform the classification over the current state-of-the-art baselines, \textit{i.e.,} Pattern-based \cite{alomar2019can} and Random classifier \cite{da2017using}. 
The selection of the two baseline approaches to compare against our approach was similar to Da Silva et al. \cite{da2017using}. We opt to choose a pattern-based approach because the methods used so far to identify refactoring commits \cite{kim2014empirical,zhangpreliminary18,Ratzinger:2008:RRS:1370750.1370759,stroggylos2007refactoring,soares2013comparing,ratzinger2005improving,murphy2012we,Mauczka2012} and analyze refactoring activity \cite{soares2013comparing} heavily rely on string matching. Other studies (\textit{e.g.,} \cite{Mauczka2012}) that focused on classifying commit messages on Swanson's categories (Corrective, Adaptive, Perfective) also used keyword-based approach.  Currently, there is no evidence on how well pattern-based approaches perform. The choice of random classifier was similar to \cite{da2017using} that assumes that the detection of self-affirmed refactoring is random.
Existing studies (\textit{cf.}, Table~\ref{sec:RelatedWork}) that have applied machine learning techniques in similar contexts (i.e., text classification) usually evaluate their approach using different classifiers. To compare their approach against others, they consider the keyword-based approach. To our knowledge, the only study that considers additional approach (\textit{i.e.,} random classifier) is the study by Da Silva et al. \cite{da2017using}. Thus, we consider keyword-based and random classifiers to compare against our approach.

Answering this research question would shed light on whether the classification of SAR is a learning problem or not. We hypothesize that if learning algorithms cannot outperform a String matching algorithm, then there is no need for proposing such a framework. The two chosen baselines, for this investigation, are listed below: 
\begin{itemize}
    \item \textbf{Baseline 1 (Pattern-based technique):}
    The pattern-based approach in identifying SAR is proposed by AlOmar et al. \cite{alomar2019can}. In their work, they identified 87 recurring patterns in SAR commit messages. We use these patterns as indicators of refactoring activities, \textit{i.e.,} if a pattern exists in a commit, it is then classified as a SAR.
    
   In order to calculate the standard statistical metrics for this baseline, 
   we use a set of 1,823 and 1,044 commit messages (\textit{cf.,} Table~\ref{Table:Instances per class (train, test)}) from the list of SAR and non-SAR commits and from each class of the multiclass classification respectively. We use them to perform a manual inspection to identify true positives (\textit{tp}), true negatives (\textit{tn}), false positives (\textit{fp}), and false negatives (\textit{fn}). True positives are cases when the pattern-based approach correctly identified SAR commits, and true negatives are commits correctly classified as without SAR. Similarly, false positives are commits classified as being SAR when they are not and finally false negatives are commits classified as without SAR when they are really SAR commits. Thus, using the \textit{tp}, \textit{tn}, \textit{fp}, and \textit{fn} values, we compute the precision, recall, and F-measure.
    
    \item \textbf{Baseline 2 (Random classifier):}
    Similar to Da Silva et al. \cite{da2017using}, we consider Random classifier as one of the baselines to compare against our approach. The rationale behind using this random classifier to hold our approach accountable for providing significantly better results in comparison with a random classification. The precision of this approach is calculated by taking the total number of SAR over the total number of commit messages for all projects. As for the recall, there is a 50\% chance that commit messages will be classified as SAR. The calculation of F-measure is explained previously in Section~\ref{sec:Approach}. 
\end{itemize}

Table~\ref{Table:Approaches comparision} and Figure~\ref{Chart:Visualization of the F1-measure for Different Approaches} present the experimental results of our approach compared with baseline 1 (Pattern-based), and baseline 2 (Random classifier). For our approach, we consider the highest F-measure score to compare against the other two baselines. Our approach provides an improvement over the comment patterns, outperforming it by 1.53 times and 1.45 times for two-class and multiclass respectively. We can see from Table~\ref{Table:Approaches comparision} that our approach outperforms the simple Random baseline by 1.84 times and by 22.14 times respectively for two-class and multiclass classifications. 

\begin{figure*}[h]
\centering 
\includegraphics[width=1.0\textwidth]{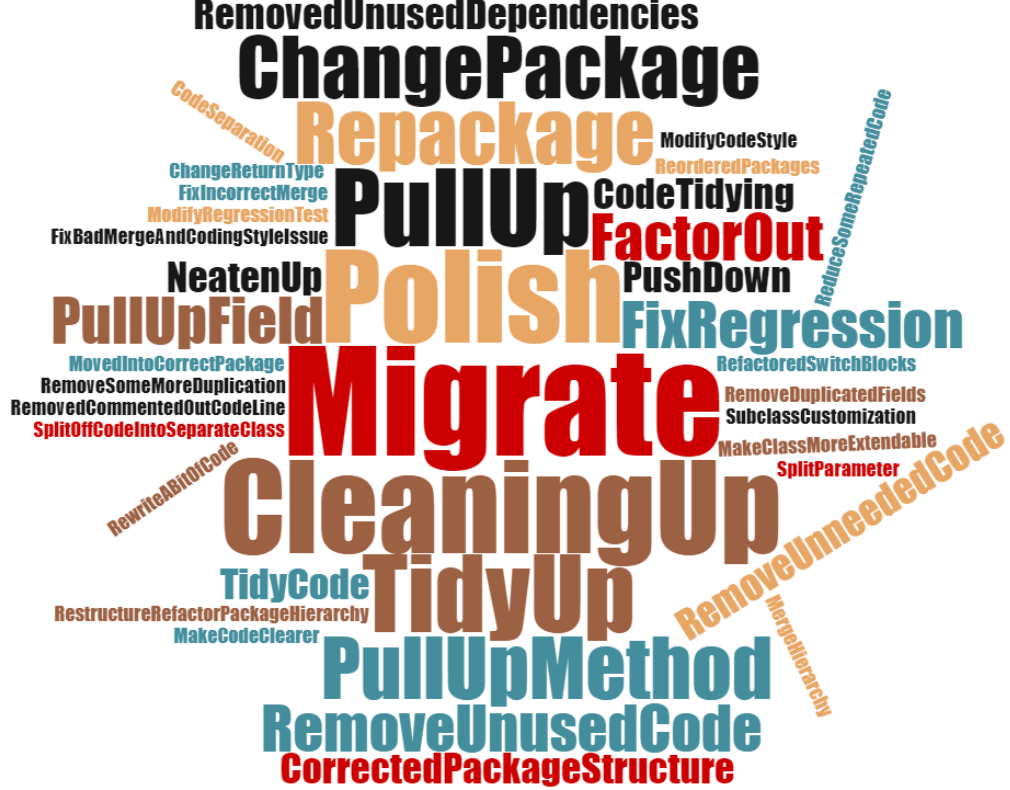}
\caption{Sample of Patterns Identified by Automatic Classification.}
\label{fig:new_patterns}
\end{figure*}

To better analyze our findings, after deploying our models as a web-service, we validate the two-class and multiclass models by randomly selecting 500 and 363 new commit messages, respectively. These new commit messages contain all types of commits (\textit{e.g.,} short commit messages, commits with more than two classes, and commits with quality attributes as part of the identifier names). We manually read through commit messages that were classified as SAR commits in the prediction phase, and were classified as non SAR according to the pattern-based approach. Intuitively, such results induce the existence of features that represent the refactoring activity, and they are not captured by the previous study. Indeed, we found a set of featured keywords that do indicate refactoring activities (\textit{e.g.,} \say{Tidy code}, \say{repackage}, and \say{fix bad merge and coding style issues}), and were not reported by any of the previous studies related to refactoring documentation. 
Such featured patterns could complement the list of manually identified 87 SAR patterns. Figure~\ref{fig:new_patterns} reveals examples of these new patterns.

\input{Tables/DifferentApproachComparison.tex}

\input{Charts/DifferentApproachVisualization.tex}

\begin{tcolorbox}
\textit{Summary}. We find that our approach can effectively outperform the classification over the current state-of-the-art baselines. We achieved an F-measure of 98\% when identifying SAR commits (an average improvement of 1.53 x and 1.84 x over the state of the art approaches), and an F-measure of 93\% when identifying the common quality improvement categories (an average improvement of 1.45 x and 22.14 x over the state of the art approaches). Additionally, our approach identifies more patterns that complement the list of manually identified 87 SAR patterns. 
\end{tcolorbox}

\subsection{\textbf{RQ3: How much training dataset is needed to effectively classify self-affirmed refactoring?}}

After assessing the accuracy of our approach in classifying SAR commits, we want to investigate the amount of training data that is needed to effectively classify SAR. Our approach will be easily extended if a small dataset can be used for SAR identification. On the other hand, if a large number of commits are required, then our approach requires considerable time and effort. 



To answer this research question, we incrementally add training data and assess the classifier’s performance. We start by randomly selecting a stratified sample of 11,000 commits for each stratum (\textit{i.e.,} SAR and non-SAR commits) provided by the authors of \cite{alomar2019can}, and combining these commits into a single large dataset. Then, we follow the classification process discussed in Section ~\ref{sec:Approach}, which results in 5,000 equally divided for each class. We then split the dataset into 10 folds with equal size, ensuring that each partition has the same ratio of SAR and non-SAR commits. For the multiclass classification problem, however, we use only a stratified sample of 1,044 commits discussed in Section~\ref{sec:Approach}. 
The reason for only considering these commits is that we are restricted by the minimum number of commits belonging to the code smell category. Thus, to avoid having an imbalanced training dataset, we keep the training size as it was originally set up. We discard the 44 commits since this number is less than the selected batch training size.   

\input{Charts/TrainingDataSize.tex}

For both classification methods, we run our approach using the 10-fold cross-validation technique, using nine folds as training data and the remaining one for testing. Because our target is to examine the impact of the quantity of training data on the performance of the classification, we train the classifier adding batches of 100 commits at a time similar to \cite{da2017using}, and evaluate their performance on the testing dataset. For each batch of commits, we maintain the same ratio of SAR and non-SAR commits. The process ends when all of the training dataset is used. After each iteration, we report the average performance for all of the folds.


Figure~\ref{Chart:TrainingSizeImpact-secondapproach-twoclass} reveals the F-measure scores when identifying SAR and non-SAR commits. Overall, we find that the F-measure maintains almost the same level with no significant improvement, in terms of accuracy, as the dataset size increases. As can be seen, we obtain a high F-measure value starting with less than 1000 commits. We conclude that only one fold of the training dataset is sufficient to identify SAR commits with F-measure of 90\%. To achieve F-measure higher than 90\%, at least one fold of 1000 is needed. 
Figure~\ref{Chart:TrainingSizeImpact-secondapproach-multiclass} shows the F-measure values when classifying Internal QA, External QA, and Code Smell commits (multiclass). In general, we notice that the F-measure value slightly increases as we increase the number of commits in the training dataset. To get at least 90\% F-measure, more than 400 commits are needed. 
We conclude that to achieve a performance equivalent to 80\% and 90\% of the high F-measure score, only 10\% and 40\% of the commit messages are required respectively. To test the significance of the difference in F-measure values, we applied the Mann-Whitney U Test and found that the differences are not statistically significant.


\begin{tcolorbox}
\textit{Summary}. We find that to achieve a performance equivalent to 90\% of the high F-measure score, only one fold of commit messages is required for the two-class and multiclass classification problems, respectively.
\end{tcolorbox}

%% file: Charts/DifferentApproachVisualization_V2.tex
\begin{figure}[tbp] 
\centering
\begin{tikzpicture}
  \centering
  \begin{axis}[
        ybar, axis on top,
        height=6cm, width=9.5cm,
        bar width=0.5cm,
        ymajorgrids, tick align=inside,
        major grid style={draw=white},
        enlarge y limits={value=.1,upper},
        ymin=0, ymax=100,
        axis x line*=bottom,
        axis y line*=right,
        y axis line style={opacity=0},
        tickwidth=0pt,
        enlarge x limits=true,
        enlarge x limits={abs=2cm}, 
        legend style={
            at={(0.5,-0.1)},
            anchor=north,
            legend columns=-1,
            /tikz/every even column/.append style={column sep=0.5cm}
        },
        ylabel={Precision (\%)},
        symbolic x coords={
           LR,RF,GBM},
       xtick=data,
       nodes near coords={
        \pgfmathprintnumber[precision=0]{\pgfplotspointmeta}
       }
    ]
    \addplot [draw=none, fill=blue!30] coordinates {
      (LR,91)
      (RF, 92)
      (GBM, 92)
      };
   \addplot [draw=none,fill=red!30] coordinates {
      (LR,92)
      (RF, 92)
      (GBM,91)
      };
   \addplot [draw=none, fill=green!30] coordinates {
      (LR, 97)
      (RF, 94)
      (GBM, 96)
       };
    \legend{Internal QA,External QA, Code Smell}
  \end{axis}
  \end{tikzpicture}
  \caption{Visualization of the Precision for Different Classifiers (Multiclass)} 

\label{Chart:Visualization of the Precision for Different Classifiers}
  \centering
\begin{tikzpicture}
  \centering
  \begin{axis}[
        ybar, axis on top,
        height=6cm, width=9.5cm,
        bar width=0.5cm,
        ymajorgrids, tick align=inside,
        major grid style={draw=white},
        enlarge y limits={value=.1,upper},
        ymin=0, ymax=100,
        axis x line*=bottom,
        axis y line*=right,
        y axis line style={opacity=0},
        tickwidth=0pt,
        enlarge x limits=true,
        enlarge x limits={abs=2cm}, 
        legend style={
            at={(0.5,-0.1)},
            anchor=north,
            legend columns=-1,
            /tikz/every even column/.append style={column sep=0.5cm}
        },
        ylabel={Recall (\%)},
        symbolic x coords={
           LR,RF,GBM},
       xtick=data,
       nodes near coords={
        \pgfmathprintnumber[precision=0]{\pgfplotspointmeta}
       }
    ]
    \addplot [draw=none, fill=blue!30] coordinates {
      (LR,93)
      (RF, 91)
      (GBM, 92)
      };
   \addplot [draw=none,fill=red!30] coordinates {
      (LR,93)
      (RF, 91)
      (GBM,92)
      };
   \addplot [draw=none, fill=green!30] coordinates {
      (LR,94)
      (RF,97)
      (GBM,95)
       };

    \legend{Internal QA,External QA, Code Smell}
  \end{axis}
  \end{tikzpicture}
  \caption{Visualization of the Recall for Different Classifiers (Multiclass).} 
\label{Chart:Visualization of the Recall for Different Classifiers}
  
  \centering

\begin{tikzpicture}
  \centering
  \begin{axis}[
        ybar, axis on top,
        height=6cm, width=9.5cm,
        bar width=0.5cm,
        ymajorgrids, tick align=inside,
        major grid style={draw=white},
        enlarge y limits={value=.1,upper},
        ymin=0, ymax=100,
        axis x line*=bottom,
        axis y line*=right,
        y axis line style={opacity=0},
        tickwidth=0pt,
        enlarge x limits=true,
        enlarge x limits={abs=2cm}, 
        legend style={
            at={(0.5,-0.1)},
            anchor=north,
            legend columns=-1,
            /tikz/every even column/.append style={column sep=0.5cm}
        },
        ylabel={F-measure (\%)},
        symbolic x coords={
           LR,RF,GBM},
       xtick=data,
       nodes near coords={
        \pgfmathprintnumber[precision=0]{\pgfplotspointmeta}
       }
    ]
    \addplot [draw=none, fill=blue!30] coordinates {
      (LR,91)
      (RF, 91)
      (GBM, 92)
      };
   \addplot [draw=none,fill=red!30] coordinates {
      (LR,92)
      (RF, 91)
      (GBM,91)
      };
   \addplot [draw=none, fill=green!30] coordinates {
      (LR,95)
      (RF,95)
      (GBM,95)
       };
    \legend{Internal QA,External QA, Code Smell}
  \end{axis}
  \end{tikzpicture}
  \caption{Visualization of the F-measure for Different Classifiers (Multiclass).} 
\label{Chart:Visualization of the F1-measure for Different Classifiers}
 \end{figure}

%% file: Tables/DifferentApproachComparison.tex
\begin{table*}[h]
\begin{center}
\caption{Comparision of Statistical Measures between our Approach, Pattern-based and the Random Classifier.}
\label{Table:Approaches comparision}
\begin{adjustbox}{width=.95\columnwidth,center}
\begin{tabular}{llllllllllll}
\toprule
\multirow{2}{*}{\textbf{Classification}}  & \multicolumn{3}{c}{\textbf{Our approach}}                                                                                & & \multicolumn{3}{c}{\textbf{Pattern-based}}                                                                             &  & \multicolumn{3}{c}{\textbf{Random Classifier}}                                                                           \\ 
& \multicolumn{1}{c}{\textbf{Precision}} & \multicolumn{1}{c}{\textbf{Recall}} & \multicolumn{1}{c}{\textbf{F-measure}} & &
\multicolumn{1}{c}{\textbf{Precision}} & \multicolumn{1}{c}{\textbf{Recall}} & \multicolumn{1}{c}{\textbf{F1}} & & \multicolumn{1}{c}{\textbf{Precision}} & \multicolumn{1}{c}{\textbf{Recall}} & \multicolumn{1}{c}{\textbf{F-measure}}  \\ 
\midrule
\textbf{Two-class} & 0.98 & 0.98 & 0.98 &                                     & 1.00 & 0.47 & 0.64 &                                   & 0.61 & 0.5 & 0.53                               \\ 
\textbf{Multiclass} & 0.93 & 0.93 & 0.93 &                                    & 0.97  & 0.48  & 0.64  &                          & 0.02  & 0.5   & 0.042          \\    
\midrule
\textbf{Two-class Improve.} & -- & -- & -- & & 0.98 x & 2.08 x & 1.53 x & & 1.60 x & 1.96 x & 1.84 x   \\
\textbf{Multiclass Improve.} & -- & -- & -- & & 0.95 x & 1.93 x & 1.45 x & &  46.5 x & 1.86 x & 22.14 x  \\
\bottomrule
\end{tabular}
\end{adjustbox}
\end{center}
\end{table*}

%% file: Charts/DifferentApproachVisualization.tex
\begin{figure}[t]  

\begin{tikzpicture}
  \centering
  \begin{axis}[
        ybar, axis on top,
        height=7cm, width=8.2cm,
        bar width=0.5cm,
        ymajorgrids, tick align=inside,
        major grid style={draw=white},
        enlarge y limits={value=.1,upper},
        ymin=0, ymax=100,
        axis x line*=bottom,
        axis y line*=right,
        y axis line style={opacity=0},
        tickwidth=0pt,
        enlarge x limits=true,
        enlarge x limits={abs=2cm}, 
        legend style={
            at={(0.5,-0.1)},
            anchor=north,
            legend columns=-1,
            /tikz/every even column/.append style={column sep=0.5cm}
        },
        ylabel={F-measure (\%)},
        symbolic x coords={
           two-class,multiclass},
       xtick=data,
       nodes near coords={
        \pgfmathprintnumber[precision=0]{\pgfplotspointmeta}
       }
    ]
    \addplot [draw=none, fill=blue!30] coordinates {
      (two-class,98)
      (multiclass, 93) 
      };
   \addplot [draw=none,fill=red!30] coordinates {
      (two-class,64)
      (multiclass, 64) 
      };
   \addplot [draw=none, fill=green!30] coordinates {
      (two-class,53)
      (multiclass, 4.2) 
       };

    \legend{Our approach,Pattern-based, Random classifier}
  \end{axis}
  \end{tikzpicture}
  \caption{Visualization of the F-measure for Different Approaches.} 
\label{Chart:Visualization of the F1-measure for Different Approaches}
  \end{figure}

%% file: Charts/TrainingDataSize.tex
\begin{figure}[h]  

  \resizebox {\columnwidth} {!} {
  \begin{tikzpicture}
     \begin{axis}[
       width=1.1\columnwidth,
       height=0.7\columnwidth,
       axis x line=bottom,
       xmin=100,
       xmax=10000,
       xlabel={Commits used in training dataset},
       xlabel near ticks,
       xticklabel style={/pgf/number format/1000 sep=},
       axis y line=left,
       ymin=0,
       ymax=100,
       scaled x ticks=false,
       ylabel={F-measure (\%)},
       ylabel near ticks, 
       legend style={
         at={(0.5,-0.23)},
            anchor=north,
    	legend columns=2,
            /tikz/every even column/.append style={column sep=0.5cm}
        },
     ]
\addplot[color=blue,mark=triangle*] coordinates {
(100, 90)
(200, 92)
(300, 94)
(400, 93)
(500, 94)
(600, 94)
(700, 95)
(800, 94)
(900, 94)
(1000, 94)
(1100, 94)
(1200, 94)
(1300, 94)
(1400, 93)
(1500, 94)
(1600, 94)
(1700, 94)
(1800, 94)
(1900, 94)
(2000,94)
(2100,95)
(2200, 95)
(2300, 95)
(2400, 96)
(2500, 96)
(2600, 95)
(2700, 95)
(2800, 95)
(2900, 95)
(3000,95)
(3100,96)
(3200, 95)
(3300, 95)
(3400, 95)
(3500, 95)
(3600, 95)
(3700, 95)
(3800, 95)
(3900, 95)
(4000,95)
(4100,95)
(4200, 95)
(4300, 95)
(4400, 95)
(4500, 95)
(4600, 95)
(4700, 95)
(4800, 95)
(4900, 95)
(5000,95)
(5100,96)
(5200, 96)
(5300, 96)
(5400, 96)
(5500, 96)
(5600, 96)
(5700, 96)
(5800, 96)
(5900, 96)
(6000,96)
(6100,96)
(6200, 96)
(6300, 96)
(6400, 96)
(6500, 96)
(6600, 96)
(6700, 96)
(6800, 96)
(6900, 96)
(7000,96)
(7100,96)
(7200, 96)
(7300, 96)
(7400, 96)
(7500, 96)
(7600, 96)
(7700, 96)
(7800, 97)
(7900, 97)
(8000,96)
(8100,97)
(8200, 97)
(8300, 97)
(8400, 97)
(8500, 97)
(8600, 97)
(8700, 97)
(8800, 97)
(8900, 97)
(9000,97)
(9100,97)
(9200, 97)
(9300, 97)
(9400, 97)
(9500, 97)
(9600, 97)
(9700, 97)
(9800, 97)
(9900, 97)
(10000,97)
       };
     \end{axis}
   \end{tikzpicture}}
 \caption{F-measure Achieved by Incrementally Adding Training Data Size for Two-class Classification.}
\label{Chart:TrainingSizeImpact-secondapproach-twoclass}

  \centering
 \end{figure}

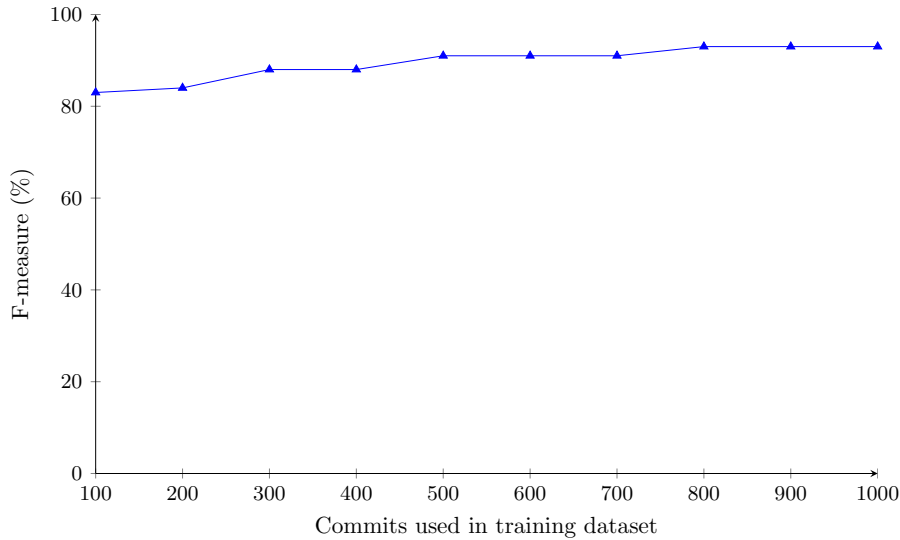
\begin{figure}[h]  
  \resizebox {\columnwidth} {!} {
  \begin{tikzpicture}
     \begin{axis}[
       width=1.1\columnwidth,
       height=0.7\columnwidth,
       axis x line=bottom,
       xmin=100,
       xmax=1000,
       xlabel={Commits used in training dataset},
       xlabel near ticks,
       xticklabel style={/pgf/number format/1000 sep=},
       axis y line=left,
       ymin=0,
       ymax=100,
       scaled x ticks=false,
       ylabel={F-measure (\%)},
       ylabel near ticks, 
       legend style={
         at={(0.5,-0.23)},
            anchor=north,
    	legend columns=2,
            /tikz/every even column/.append style={column sep=0.5cm}
        },
     ]
\addplot[color=blue,mark=triangle*] coordinates {
(100, 83)
(200, 84)
(300, 88)
(400, 88)
(500, 91)
(600, 91)
(700, 91)
(800, 93)
(900, 93)
(1000,93)
       };
     \end{axis}
   \end{tikzpicture}}
\caption{F-measure Achieved by Incrementally Adding Training Data Size for Multiclass Classification.}
\label{Chart:TrainingSizeImpact-secondapproach-multiclass}
 \end{figure}

%% file: Sections/Implications.tex
\section{Research Implications}
\label{sec:Implications}

This section further discusses positions our work in the spectrum of existing studies and how it implicates current research and practice.

\subsection{Implications for practitioners}

From a practitioner's point of view, giving enough background related to the performed refactorings is important to facilitate the code review process. Since there is no consensus on how to formally document refactoring activities, our model can provide various examples of how refactoring activity has been documented. Such information can be valuable to provide examples either to learn from or criticize. Also, since documenting code changes is enforced practice for some companies, then our tool can be used, in synchrony with other refactoring miners to detect when a refactoring, in the source code level, has no \say{expected} documentation in the commit message level. Such a quick sanity check can remind developers of adding any missing information. Furthermore, the review process heavily relies on understanding the context of the preformed refactoring, and since refactoring impact cannot be narrowed into one category, authors have to clearly state their intention in order for the reviewers to properly assess it.

Further, understanding maintenance activities is critical for practitioners to effectively direct the evolution of their projects in terms of enhancing cost-effectiveness, managing technical debt, and better planification of maintenance related resources. Therefore, a plethora of studies have been performed on automatic classification of repository artifacts (\textit{e.g.,} bug reports, issues, code changes) in general, and commit messages in particular for several purposes, including the approximation of maintenance activity \cite{gharbi2019classification,Levin:2017:BAC:3127005.3127016,honel2019importance}, security-relevant changes \cite{rosen2015commit,eyolfson2014correlations}, bug proneness \cite{eyolfson2014correlations,eyolfson2011time}, bug fixes \cite{zafar2019towards,Sadiq2019onthe}. Our work extends this existing effort by adding another dimension of the localization of refactoring effort. The end goal of estimating maintenance activities is to support managers and developers in better evaluating the quality of their projects, and so being more sensitive to anomalies that may arise, and the way to cope with them. 

These three categories provide software practitioners with a catalog of common refactoring documentation patterns which represent concrete examples of common ways to document refactoring activities in commit messages. Having these higher-level categories helps developers find the specific refactoring patterns they are looking for faster. Generally, in industry, there is no guideline on how to structure commit messages. This catalog of SAR patterns can encourage developers to follow best documentation patterns and also to further extend these patterns to improve refactoring documentation in particular and code changes in general. This work will also help developers to improve the quality of the refactoring documentation and trigger the need to explore the motivation behind refactoring. Further, these categories tell the opinion of developers, so it is important for managers to learn developers' opinions and feelings especially for distributed software development practices. If developers did not document, managers will not know their intention. Since software engineering is a human-centric process, it is important for managers to understand the intention of people working on the team.  In this work, we (1) learn  about how people self-report their types of work to evaluate progress with respect to goals for improving code quality, and (2) examine changes over time in how developers report their own activity in order to gain insight into patterns/find areas for improvement.

Moreover, for refactoring recommendations, if we know the intention of developers (\textit{e.g.,} fix code smell), we can recommend refactoring based on the intention. From refactoring commit messages, we learn from these commit message examples and know what code elements they change, we then can optimize our refactoring recommendation to just work on code elements they are changing. This work will help refactoring recommending systems by narrowing their scope (\textit{e.g.,} working on code fragments that developers are interested in). Current recommender system did not look at the intention, they excluded completely the intention of developers. Thus, these recommender systems are underused because they did not consider this important aspect.

\subsection{Implications for researchers}

From a research perspective, recent studies have been focusing on automatically identifying any execution of a refactoring operation in the source code \cite{silva2017refdiff,tsantalis2018accurate,kim2010ref}. The main purpose of the automatic detection of refactoring is to better understand how developers cope with their software decay by extracting any refactoring strategies that can be associated with removing code smells \cite{tsantalis2008jdeodorant,bavota2013empirical}, or improving the design structural measurements \cite{bavota2014recommending,mkaouer2014recommendation}. However, these techniques only analyze the changes at the source code level, and provide the operations performed, without associating it with any textual description, which may infer the rationale behind the refactoring application. Our proposed model intends to bridge this gap by complementing the existing effort in accurately detecting refactorings, by augmenting with any description that was intended to describe the refactoring activity. As previously shown in Tables \ref{Table:Patterns} and \ref{Table:QA & Code smell}, developers tend to add a high-level description of their refactoring activity, and occasionally mention their intention behind refactoring (remove duplicate code, improve readability), along with mentioning the refactoring operations they apply (type migration, inline methods, etc.). Our model, combined with the detection of refactoring operations, serves as a solid background for various empirical investigations. For instance, previous studies have analyzed the impact of refactoring operations on structural metrics \cite{palomba2017exploratory,bavota2015experimental,cedrim2016does}. One of the main limitations of these studies is the absence of any context related to the application of refactorings, \textit{i.e.,} it is not clear whether developers did apply these refactoring with the intention of improving design metrics. Therefore, the use of our model will allow the consideration of commits whose commit messages specifically express the refactoring for the purpose of optimizing structural metrics, such as coupling, and complexity, and so, many empirical studies can be revisited with a more adequate dataset. 

Furthermore, our study provides software practitioners with a catalog of common refactoring documentation patterns (cf. Tables \ref{Table:Patterns} and \ref{Table:QA & Code smell}) which would represent concrete examples of common ways to document refactoring activities in commit messages. This catalog of SAR patterns can encourage developers follow best documentation patterns and also to further extend these patterns to improve refactoring changes documentation in particular and code changes in general. Indeed, reliable and accurate documentation is always of crucial importance in any software project. The presence of documentation for low level changes such as refactoring operations and commit changes helps to keep track of all aspects of software development and it improves on the quality of the end product. Its main focuses are learning and knowledge transfer to other developers.

Another important research direction that requires further attention concerns the documentation of refactoring. It has been known that there is a general shortage of refactoring documentation, as developers typically focus on describing their functional updates and bug patches. Also, there is no consensus about how refactoring should be documented, which makes it subjective and developer specific. Moreover, the fine-grained description of refactoring can be time consuming, as typical description should contain indication about the operations performed, refactored code elements, and a hint about the intention behind the refactoring. In addition, the developer specification can be ambiguous as it reflects the developer's understanding of what has been improved in the source code, which can be different in reality, as the developer may not necessarily adequately estimate the refactoring impact on the quality improvement. Therefore, our model can help to build a corpus of refactoring descriptions, and so many studies can better analyze the typical syntax used by developers in order to develop better natural language models to improve it, and potentially automate it, just like existing studies related to other types of code changes \cite{buse2010automatically,linares2015changescribe,liu2018neural}.

This work can help researchers to investigate the consistency between code changes and the actual intention and explore whether there is an overlap or not.

\subsection{Implications for educators}

From an educator point of view, this study helps to teach the new generation of developers or engineers the best practice to document their refactoring activity.

%% file: Sections/Threats.tex
\section{Threats to Validity}
\label{sec:Threats}

In this section, we identify potential threats to the validity of our approach and our experiments. 

\begin{figure*}[h]
\centering 
\includegraphics[width=0.9\linewidth]{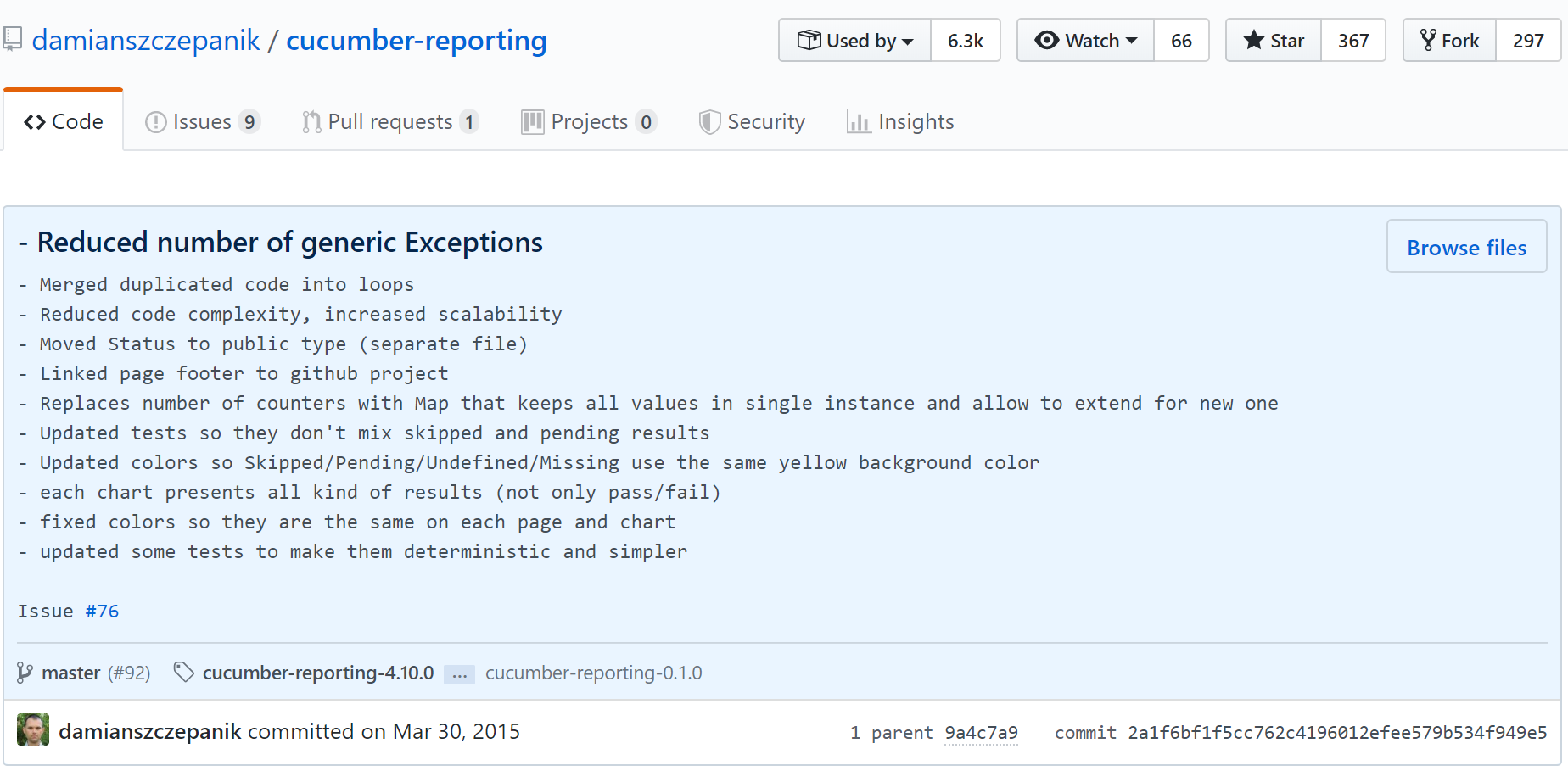}
\caption{Example of Multi-label Refactoring Commit Message that Would Confuse the Classifiers.}
\label{fig:github-multiclass-future}
\end{figure*}

\textbf{Construct Validity:} Since our approach heavily depends on commit messages, we used well-commented Java projects when performing our study. Thus, the quality and the quantity of commit messages might have an impact on our findings. Additionally, a well-commented project might not contain SAR as developers might not document refactoring activities in the commit messages. We mitigate this risk by choosing projects that are appropriate for our analysis. Another potential threat relates to manual classification. Since the manual classification of training commit messages is a human intensive task and it is subject to personal bias, we mitigate manual classification related errors by discarding short and ambiguous commits from our dataset and replacing them with other commits. Another important limitation concerns the size of the dataset used for training and evaluation. The size of the used dataset was determined similarly to previous commit classification studies, but we are not certain that this number is optimal for our problem. It is better to use a systematic technique for choosing the size of the evaluation set. Concerning the relationship between refactoring and quality issues, we designed our study with the goal of classifying refactoring documentation. We have not explored if the refactoring operations detected by the Refactoring Miner tool are related to the corresponding quality issues documented by developers  in the commit messages. Further, recent studies \cite{YAN2016296,Kirinuki:2014:HYC:2597008.2597798,doi:10.1287/isre.9.2.101} indicate that commit messages could capture more than one type of classification (\textit{i.e.,} mixed maintenance activity). Figure~\ref{fig:github-multiclass-future} shows a commit message could belong to internal quality attribute (since it discusses code complexity reduction), external quality attribute (since it points out scalability improvement), and code smell (since it explains duplicate code removal). In this work, we have not yet investigated whether a significant number of commits can belong to more than once class, and if so, we plan on exploring a multi-label classification in our future work.

\textbf{External Validity:} The first threat relates to the commits that are extracted only from open source Java projects. Our results may not generalize to commercially developed projects, or to other projects using different programming languages. Another threat concerns the generalization of SAR patterns in the commit messages. Since a commit is considered SAR commits if it only contains any of SAR patterns, this may not generalize to other projects (\textit{e.g.,} outside the Java developers community) as it may have additional expression that could belong to SAR category.  

Although we used commit messages as our primary source of text, our approach is not restricted to a specific source of textual information. In our future work, we can replicate our approach with other types of metadata, including issue descriptions. For this study, we chose to focus on commit messages rather than issue descriptions, since issue descriptions can be very high level, may not go into code change details, may not always be available, and may refer to multiple changes in the code that span or mix different purposes (\textit{e.g.,} bug fix and feature request). Besides, not all projects are using issue tracker. If the issue tracker is guaranteed to be available, it could be used as an additional source of information.

The use of well engineered projects is a double-edged sword, while it guarantees an easier labeling process, and providing less noisy data for the approach, it hinders its generalizability since these projects represent only a subset of all projects. So, our model may not achieve similar (high) performance across many projects. We tried to mitigate this concern by considering different types of projects, belonging to different domains. We shuffled commit messages during the training and testing to avoid any biases.

%% file: Sections/Conclusion.tex
\section{Conclusion}
\label{sec:conclusion}
In this paper, we proposed an approach to identify and classify self-affirmed refactoring in commit messages. We compared the performance of our approach to pattern-based and simple random baselines. Our results show that our approach  (1) is able to accurately classify SAR commits with accuracy of 98\% and 93\% for two-class and multiclass classification methods, respectively, outperforming the two state-of-the-art approaches considered in this study, and (2) can achieve F-measure of 90\% using only 1\% and 40\% of the commits when performing two-class and multiclass SAR classifications respectively. This indicates a relatively small training dataset is sufficient to classify SAR commits. 

In the future, we plan to study the applicability of our approach to other projects developed in different programming languages, and to other domains. Another potential research direction is to use the current findings to build a tool that supports the identification and detection of self-affirmed refactoring commits. We also plan to conduct different user studies with our industrial partner predict the refactoring intention of the developers and further assess whether it aligns with what happened to his source code after applying refactoring.